\documentclass[iop]{emulateapj}

\usepackage{hyperref}
\slugcomment{To appear in The Astrophysical Journal}

\newcommand{\teff}{T_\mathrm{eff}}
\newcommand{\logg}{\log g}
\newcommand{\feh}{\mathrm{[Fe/H]}}
\newcommand{\fei}{Fe \textsc{I}}
\newcommand{\hei}{He \textsc{I}}
\newcommand{\sii}{Si \textsc{I}}
\newcommand{\feii}{Fe \textsc{II}}
\newcommand{\vt}{v_t}
\newcommand{\nan}{\nodata}
\newcommand{\nai}{Na \textsc{I}}
\newcommand{\ali}{Al \textsc{I}}
\newcommand{\mgi}{Mg \textsc{I}}
\newcommand{\oi}{O \textsc{I}}
\newcommand{\baii}{Ba \textsc{II}}
\newcommand{\cai}{Ca \textsc{I}}
\newcommand{\cri}{Cr \textsc{I}}
\newcommand{\tii}{Ti \textsc{I}}
\newcommand{\si}{S \textsc{I}}

\begin{document}

\title{The Kapteyn moving group is not tidal debris from $\omega$~Centauri\footnotemark[1]}

\footnotetext[1]{Based on observations collected at the European Southern Observatory, Chile 
(ESO Program 090.B-0605) and observations gathered with the 6.5 m Magellan Telescopes at Las Campanas Observatory, Chile.}

\author{Camila Navarrete\altaffilmark{2,}\altaffilmark{3}, 
Julio Chanam\'e\altaffilmark{2,}\altaffilmark{3}, 
Iv\'an Ram\'irez\altaffilmark{4}, 
Andr\'es Meza\altaffilmark{5},
Guillem Anglada-Escud\'e\altaffilmark{6},
Evgenya Shkolnik\altaffilmark{7}
}

\altaffiltext{2}{Instituto de Astrof\'isica, Pontificia Universidad Cat\'olica de Chile, 
Av. Vicu{\~n}a Mackenna 4860, 782-0436 Macul, Santiago, Chile. \email{cnavarre@astro.puc.cl}}

\altaffiltext{3}{Millennium Institute of Astrophysics, Santiago, Chile.}

\altaffiltext{4}{Department of Astronomy, University of Texas at Austin, 2515 Speedway, 
Stop C1402, Austin, TX 78712-1206, USA.}

\altaffiltext{5}{Departamento de Ciencias F\'isicas, Universidad Andr{\'e}s Bello, 
Av. Rep\'ublica 220, Santiago, Chile.}

\altaffiltext{6}{Astronomy Unit, School of Mathematical Sciences, Queen Mary, University of London, London E1 4NS, UK}

\altaffiltext{7}{Lowell Observatory, 1400 W. Mars Hill Road, Flagstaff, AZ, USA.}

\renewcommand{\thefootnote}{\fnsymbol{footnote}}

\begin{abstract}
The Kapteyn moving group has been postulated as tidal debris from $\omega$~Centauri. If true, members 
of the group should show some of the chemical abundance patterns known for stars in the cluster. 
We present an optical and near-infrared high-resolution, high-signal-to-noise ratio spectroscopic study of 14 stars 
of the Kapteyn group, plus 10 additional stars (the $\omega$~Cen group) that, while not listed as 
members of the Kapteyn group as originally defined, have nevertheless been associated dynamically 
with $\omega$~Centauri. Abundances for Na, O, Mg, Al, Ca, and Ba were derived from the optical spectra, 
while the strength of the chromospheric \hei\ 10830 {\AA} line is studied as a possible helium 
abundance indicator. The resulting Na--O and Mg--Al patterns for stars of the combined Kapteyn and 
$\omega$~Cen group samples do not resemble those of $\omega$~Centauri, and are not different from 
those of field stars of the Galactic halo. The distribution of equivalent widths of the \hei\ 10830 
{\AA} line is consistent with that found among non-active field stars. Therefore, no evidence is 
found for second-generation stars within our samples, which most likely rules out a globular-cluster 
origin. Moreover, no hint of the unique barium overabundance at the metal-rich end, well established 
for $\omega$~Centauri stars, is seen among stars of the combined samples. Because this specific Ba 
pattern is present in $\omega$~Centauri irrespective of stellar generation, this would rule out 
the possibility that our entire sample might be composed of only first-generation stars from the 
cluster. Finally, for the stars of the Kapteyn group, the possibility of an origin in the hypothetical 
parent galaxy of $\omega$~Centauri is disfavored by the different run of $\alpha$-elements with metallicity 
between our targets and stars from present-day dwarf galaxies.
\end{abstract}


\renewcommand{\thefootnote}{\arabic{footnote}}
\setcounter{footnote}{0}

\section{Introduction}
\label{sec:intro}

Stellar halos of galaxies such as the Milky Way (MW) are thought to form hierarchically via the 
accumulation of stars from smaller satellite galaxies formed at earlier times \citep{S78, F02}. 
Strong evidence for this scenario has been found in the MW as well as in extragalactic 
environments, e.g., in the form of accreting galaxies and star clusters, often displaying 
long tidal tails and organized streams of stars \citep{B06a, MD12}. Indeed, the accretion 
of independent stellar systems into the MW is taking place even today, notably demonstrated by 
the Sagittarius stellar stream \citep{I94, I95, V01, M03} and many others.
 
These disrupting systems are depositing their stellar populations into the Galactic field, 
particularly the stellar halo, and therefore the study of the detailed chemistry and kinematics 
of halo stars, and their comparison with stars in the aforementioned stellar streams, should shed 
light on the processes and the progenitors responsible for this hierarchical mode of MW assembly. 
Achieving this, however, is complicated by the large distances to the presently occurring, easily 
identifiable, accretion events, which prevent us from studying them in all the detail that is possible 
when studying more local populations. Therefore, it would be desirable to study instead the 
remnants of accretion events existing today within the solar neighborhood. However, identifying 
local stars that have been accreted throughout the MW's history is not an easy task because most 
of these should have mixed by now with the background Galactic stellar populations.

Despite such difficulties, efforts to identify accreted stars in the Galactic field have already 
started to uncover what appears to be the long sought-after population of halo field stars that 
originated in disrupted globular clusters (GCs), particularly those corresponding to second-generation 
stars\footnote{Hereafter, the term ``generations'' will be used to refer to groups of GC 
stars that present differences in their light-element abundances, such as Na, O, Mg, and Al, as well 
as in their helium content.}. First, \cite{Martell11} found 3\% of red giant stars with CN and CH 
bandstrengths consistent with depleted carbon and enhanced nitrogen abundances, an atypical light-element 
pattern found only in GC second-generation stars. The same fraction was obtained by \cite{R12}, who 
discovered the first examples of field dwarf stars showing low oxygen and high sodium abundances, which 
are key signatures of the O--Na anticorrelation, a characteristic abundance pattern seen only 
among GC stars. Based on the models of \cite{V10} for the evolution of GCs in the field of the 
Galaxy, these results suggest to us that a fraction as large as 40-50\% of the present Galactic 
stellar halo might be composed of stars, from both first and second generations, originating in such 
stellar systems.

Another opportunity for identifying field stars originally belonging to GCs might be provided by 
the measurement of helium abundances. Although initially unpopular, the hypothesis of stellar 
generations with high helium abundance \citep[up to $Y\sim$~0.40,][]{N04} arguably remains as the 
most solid explanation for the observed splitting of the main sequences (MSs) of $\omega$~Centauri 
and various other GCs, while at the same time opening possibilities for the natural resolution of 
additional long-standing puzzles. In fact, spectroscopic studies of stars in the different MSs of 
some GCs found that the blue MS is more metal-rich than the red one, something that could be 
explained only if the blue MS has a higher He abundance \citep{D05, P05, P07}. Recent spectroscopic 
studies that could directly measure the He abundance in the horizontal branch of GCs known to host 
different MSs confirmed this scenario, and also found that wider MS splitting requires higher helium 
enhancement \citep[see e.g.,][]{V12, M14, Mucciarelli14, V14, M15}. Previous to these recent 
developments, there has been a long history of well documented abundance anomalies as tracked by 
the luminous red giants in GCs, and extended down to below the MS turn-off in more recent years. 
Notorious among these are correlated and anticorrelated abundances of elements such as C, N, O, 
Na, Mg, Al, and others, all thought to arise from He-producing, high-temperature branches of 
hydrogen-burning chains \citep[see e.g.,][and references therein]{G04, C07, C09, C09b, Carretta10c, G12}. 

In this context, first-generation GC stars are those that have chemical abundances indistinguishable 
from the MW halo field stars, at a given metallicity. Second-generation GC stars present ``anomalies'' 
in the chemical abundances such as low oxygen, magnesium, and carbon, and enhanced sodium, nitrogen, and 
aluminum \citep{C09, C09b, Martell11}. Spectroscopic studies confirmed that second-generation GC stars 
are also helium-enhanced \citep{P11, D13, M14}.

Most of the uncommon characteristics of the $\omega$~Centauri GC, such as its wide metallicity 
distribution, abundance anomalies \citep{N95}, multiple stellar populations\footnote{By ``populations'' we 
are referring to different mean metallicity peaks in the metallicity distribution of a group of stars.} 
\citep[]{B04}, mildly retrograde orbit \citep{D99}, among others, strongly suggest that it may 
be the remnant nucleus of an accreted dwarf galaxy from a past merger event \citep{F93, B03, L09}. 
This might make $\omega$~Centauri the nearest surviving remnant of a true Galactic building block, 
one from which stars were stripped and incorporated into the Galactic field populations. Thus it would 
be of significant interest to identify stars in the solar neighborhood that could be unambiguously 
linked to this merger event.

Chemical abundances measured for the different populations observed along the red giant branch (RGB) of $\omega$~Centauri 
have shown that the cluster had a complex chemical enrichment history. In fact, despite the cluster 
being suspected to be the remnant nucleus of a disrupted dwarf galaxy, it has [$\alpha$/Fe] ratios 
between 0.0 and 0.26 dex, more similar to the MW field stars \citep{G11} than to dwarf galaxies, which 
display sub-solar [$\alpha$/Fe] abundances \citep[see][]{T09}. At the same time, $\omega$~Centauri 
has higher s-process abundances, as other dwarf galaxies show \citep{T09}, but the fast rise in the 
barium abundance with metallicity seen in stars in $\omega$~Centauri is not found in any other 
galaxy or star cluster \citep[see, e.g.,][]{G07, JP10, M11}. Such specific variation of Ba abundances with 
metallicity among stars in $\omega$~Centauri is extremely important because it is independent of 
whether stars are from the first or second generation, and it is displayed by all populations in the 
cluster. Besides the overabundance of s-process elements, $\omega$~Centauri displays the well-known 
O--Na and Mg--Al anticorrelations found in most GCs \citep{JP10}. The O--Na anticorrelation 
is found over almost the whole metallicity range measured in the cluster, except in the most metal-rich 
population \citep{G11, M11}. The anticorrelation is more marked than that observed typically in GCs, 
but similar to NGC~2808, another of the most massive GCs in the MW \citep{G11}. Like other GCs, 
$\omega$~Centauri has almost equal fractions of first- and second-generation stars in the metal-poor 
population, but an exceptionally high fraction of second-generation stars in the metal-intermediate 
and metal-rich populations \citep[less than 20\% of the stars associated with the intermediate-metallicity 
population have abundances consistent with first-generation stars; see][and references therein]{JP10}.

The Kapteyn moving group was originally identified as composed of stars seemingly distinct, 
kinematically speaking, from the bulk disk and halo local populations \citep{E78, E96, NHF04}. 
Recently, \citet[WdB10 hereafter]{W10} studied chemical abundances of some members of the Kapteyn 
group, concluding that they have abundance patterns similar to those found among $\omega$~Centauri 
red giants, including abundance ratios involving Na, Ba, and s-process elements that are enhanced 
compared to the field halo stars. Furthermore, they argued that stars in the Kapteyn group are 
nicely located in the range of expected values of angular momentum and energy for $\omega$~Centauri  
tidal debris \citep{D02, M05}. Based on those arguments, \defcitealias{W10}{WdB10} \citetalias{W10} suggested that some stars in the 
Kapteyn group may be part of the tidal debris from the same merger event that disrupted  
$\omega$~Centauri's parent dwarf galaxy. However, errors in the abundance determinations of \citetalias{W10}, 
added to the use of literature values from heterogeneous sources to compare the populations of the 
halo and $\omega$~Centauri with the Kapteyn stars, justify a re-examination of this conclusion. 
Furthermore, \citetalias{W10} do not consider the oxygen abundance, which follows a very marked anticorrelation 
with sodium in GC stars and thus could constitute a key ingredient to properly disentangle between 
field and GC origins \citep{C09}. Hence, the association between the Kapteyn moving group and 
$\omega$~Centauri stars should be considered with caution and not as unambiguously established.

Moreover, as discussed earlier, another species that could be used to uncover field stars of GC 
origins is helium, although using this element for the chemical tagging of field stars has not 
been attempted yet, to the best of our knowledge. Measuring the abundance of helium is a significant 
observational challenge because He absorption lines need quite high temperatures to be formed 
(higher than $\sim$8000 K), and even at these temperatures the optical \hei\ line 
($\lambda$~5875.5 {\AA}) is weak and needs high signal-to-noise ratio (S/N) spectra to be properly measured 
\citep{V09,V12}. Additionally, at higher temperatures ($T_{\rm eff} \gtrsim$~11500~K), helium 
suffers of sedimentation and settling, diminishing the strength of the absorption line and, hence, 
yielding underestimated photospheric He abundances. Cool stars are not hot enough to 
produce the He transition in the optical, but they commonly display the near-IR \hei\ 10830 
{\AA} chromospheric line. This line historically has been used as a chromospheric activity indicator 
for late-type giant and supergiant stars \citep{O86, T11, S12}; and as a sensitive probe of mass 
loss by stellar winds \citep{D09}. Moreover, \cite{S08} have demonstrated a direct relation between 
the \hei\ line and the extreme ultraviolet and X-ray emission flux related to coronal activity in 
cool stars.

The connection between the chromospheric He line and the possibility of helium enhancement has been 
scarcely studied but the few available results are quite suggestive. \cite{D11} studied red giants 
in $\omega$~Centauri using infrared and optical spectra, finding that the stars that have larger \hei\ 
10830 {\AA} equivalent widths (EWs) also have the largest [Al/Fe] and [Na/Fe] abundances, arguably deriving 
the first convincing evidence for an enhancement of helium in giant stars in this cluster. 
A similar study was carried out by \cite{P11}, who studied the difference between two giants 
of NGC 2808 with similar stellar parameters but strongly different Na and O abundances, finding 
that a helium enhancement of $\Delta Y \geq 0.17$ in the Na-rich, O-poor star is required to fit 
theoretical chromospheric models in their infrared spectra. Recently, \cite{D13} used 
state-of-the-art radiative transfer models considering a semi-empirical non-LTE spherical atmosphere, 
in addition to a chromospheric model, deriving the helium abundance of two giants from $\omega$~Centauri; 
they found that the star that is Na- and Al-rich is consistent with $Y = 0.39-0.44$ with a marked 
chromospheric \hei\ line in absorption. This is, to our knowledge, the only work that derives a 
helium abundance based on the \hei\ 10830 {\AA} line.

In this paper we aim to study the chemistry and kinematics of the Kapteyn moving group in order to 
establish whether it can be unambiguously associated to $\omega$~Centauri or not. Moreover, we add 
to our sample a number of stars from the kinematically selected ``$\omega$~Cen group'', as defined 
by \cite{M05} using the angular momentum distribution of metal-poor field stars. If both groups 
are constituted by stripped stars from $\omega$~Centauri, we expect at least four and three stars with 
second-generation chemical abundances in our sample, respectively. Moreover, they also should present 
the unique Ba--$\feh$ pattern, which is independent of the stellar generation. From the chemical 
analysis, the classical correlations and anticorrelations expected in GC stars are studied. 
At the same time, the possibility of He enhancement is addressed using the near-IR \hei\ 
chromospheric line. The paper is organized as follows. In Section~\ref{sec:datos} the observations and data 
reduction from the optical and infrared spectra are presented. The measurement of stellar 
parameters for the full sample and the determination of elemental abundances and near-IR \hei\ 
EWs are described in Section~\ref{sec:abundances}. The resulting chemical patterns are discussed in 
Section~\ref{sec:chemical}, and in Section~\ref{sec:kine} we briefly examine the kinematics of the studied samples. 
Our conclusions are presented in Section~\ref{sec:conclusions}.

\section{Observations and data reduction}
\label{sec:datos}

\subsection{Target Selection}

We target stars from the Kapteyn group, an apparently coherent kinematic structure within the solar 
neighborhood that has recently been linked, both kinematically and chemically, to the GC
$\omega$~Centauri \citepalias{W10}. \cite{E96} listed 33 stars kinematically associated to the Kapteyn 
group \citep[earlier defined by][]{E77}, from which there are 7 that are variable. Our sample 
collects 14 of the remaining 26 non-variable stars, all of them included in the study of \citetalias{W10}, 
which contains a total of 16 stars, extracted from the study of \cite{E96}.

We also include in our sample stars from the ``$\omega$~Cen group'' of \cite{M05}, identified as 
having specific angular momenta similar to that of the $\omega$~Centauri cluster and slightly retrograde 
orbits ($\mathbf{-}$50 km s${}^{-1}$ $<$ V${}_{\rm rot}$ $<$ 0 km s${}^{-1}$). These stars are part of the 
catalogs of metal-poor field stars of \cite{B00} and \cite{G03}. 

A comparison sample of field stars was also observed. These stars were selected from the catalog of \cite{G03},  
such that they span the range of stellar parameters of stars in the Kapteyn and $\omega$~Cen groups.

Table~\ref{targets} lists our target stars with their fundamental atmospheric parameters, as determined below.

\begin{deluxetable*}{lrllr}
\tablecolumns{5}
\tablecaption{Spectroscopic stellar parameters of our target stars.\label{targets}}
\tablehead{\colhead{Star} & \colhead{$V{}_{\rm hel}$} & \colhead{$T{}_{\rm eff}$} & \colhead{$\log{g}$} & \colhead{[Fe/H]}  \\ 
\colhead{} & \colhead{(km s${}^{-1}$)} & \colhead{(K)} & \colhead{(dex)} & \colhead{(dex)}  } 
\startdata
\multicolumn{5}{c}{\textbf{Kapteyn Group Stars}}\\
\hline
CD-30 1121                    &  105.25 $\pm$ 0.33 & 4955 $\pm$  65 & 2.07 $\pm$ 0.16 & -1.919 $\pm$ 0.06   \\ 
CD-62 1346                    &  125.74 $\pm$ 0.35 & 5524 $\pm$  83 & 2.16 $\pm$ 0.18 & -1.399 $\pm$ 0.06   \\ 
HD 110621                     &  222.17 $\pm$ 0.33 & 6212 $\pm$  97 & 4.04 $\pm$ 0.20 & -1.522 $\pm$ 0.07   \\ 
HD 111721\tablenotemark{a}  &   21.53 $\pm$ 0.35 & 4990           & 2.60            & -1.380               \\ 
HD 13979                      &   52.38 $\pm$ 0.36 & 5023 $\pm$ 145 & 1.47 $\pm$ 0.22 & -2.441 $\pm$ 0.08    \\ 
HD 181007                     &   -1.72 $\pm$ 0.35 & 4979 $\pm$  51 & 2.23 $\pm$ 0.13 & -1.721 $\pm$ 0.05    \\ 
HD 181743                     &   29.50 $\pm$ 0.37 & 5998 $\pm$ 123 & 4.26 $\pm$ 0.26 & -1.795 $\pm$ 0.09    \\ 
HD 186478                     &   31.45 $\pm$ 0.35 & 4568 $\pm$  85 & 0.88 $\pm$ 0.19 & -2.500 $\pm$ 0.07    \\ 
HD 188031                     & -144.21 $\pm$ 0.40 & 6288 $\pm$  81 & 4.24 $\pm$ 0.16 & -1.649 $\pm$ 0.06    \\ 
HD 193242                     & -127.85 $\pm$ 0.39 & 5031 $\pm$  54 & 2.35 $\pm$ 0.14 & -1.778 $\pm$ 0.05    \\ 
HD 208069                     & -166.59 $\pm$ 0.41 & 5065 $\pm$  46 & 2.47 $\pm$ 0.12 & -1.741 $\pm$ 0.04    \\ 
HD 21022                      &  122.85 $\pm$ 0.33 & 4524 $\pm$  62 & 0.89 $\pm$ 0.15 & -2.204 $\pm$ 0.05    \\ 
HD 215601                     &  -36.65 $\pm$ 0.36 & 4972 $\pm$  45 & 1.67 $\pm$ 0.12 & -1.460 $\pm$ 0.04    \\ 
HD 215801                     &  -90.87 $\pm$ 0.40 & 6065 $\pm$ 184 & 3.69 $\pm$ 0.18 & -2.183 $\pm$ 0.06    \\ 
\hline
\multicolumn{5}{c}{\textbf{$\omega$~Cen group Stars}}\\
\hline
BD+02 3375                    & -397.77 $\pm$ 0.49 & 6009 $\pm$ 160 & 4.01 $\pm$ 0.18 & -2.190 $\pm$ 0.06    \\ 
CD-61 0282                    &  220.35 $\pm$ 0.34 & 5815 $\pm$  70 & 4.30 $\pm$ 0.16 & -1.234 $\pm$ 0.06    \\ 
HD 113083\tablenotemark{a}  &  227.91 $\pm$ 0.46 & 5800           & 4.35            & -0.930               \\ 
HD 121004                     &  245.63 $\pm$ 0.33 & 5651 $\pm$  48 & 4.31 $\pm$ 0.11 & -0.784 $\pm$ 0.04    \\ 
HD 140283                     & -170.36 $\pm$ 0.39 & 5788 $\pm$  79 & 3.56 $\pm$ 0.13 & -2.393 $\pm$ 0.05    \\ 
HD 148816                     &  -47.62 $\pm$ 0.34 & 5876 $\pm$  43 & 4.16 $\pm$ 0.11 & -0.790 $\pm$ 0.04    \\ 
HD 193901                     & -171.39 $\pm$ 0.39 & 5721 $\pm$  44 & 4.42 $\pm$ 0.11 & -1.109 $\pm$ 0.04    \\ 
HD 194598                     & -247.04 $\pm$ 0.40 & 5998 $\pm$  47 & 4.33 $\pm$ 0.12 & -1.148 $\pm$ 0.04    \\ 
HD 3567                       &  -47.60 $\pm$ 0.35 & 6130 $\pm$  51 & 4.06 $\pm$ 0.12 & -1.200 $\pm$ 0.04    \\ 
HD 84937                      &  -15.12 $\pm$ 0.39 & 6280 $\pm$ 150 & 4.00 $\pm$ 0.20 & -2.089 $\pm$ 0.07    \\ 
\hline
\multicolumn{5}{c}{\textbf{Field Stars}}\\
\hline
HD 102200                     &  161.26 $\pm$ 0.33  & 6113 $\pm$  62 & 4.15 $\pm$ 0.15 & -1.222 $\pm$ 0.05    \\ 
HD 116064                     &  140.26 $\pm$ 0.38  & 6010 $\pm$ 195 & 4.16 $\pm$ 0.22 & -1.848 $\pm$ 0.07    \\ 
HD 128279                     &  -75.44 $\pm$ 0.35  & 5245 $\pm$  97 & 2.79 $\pm$ 0.23 & -2.123 $\pm$ 0.08    \\ 
HD 134439                     &  310.04 $\pm$ 0.34  & 4956 $\pm$  69 & 4.60 $\pm$ 0.17 & -1.440 $\pm$ 0.06    \\ 
HD 134440                     &  310.77 $\pm$ 0.35  & 4795 $\pm$  66 & 4.60 $\pm$ 0.16 & -1.426 $\pm$ 0.06    \\ 
HD 142948                     &   29.28 $\pm$ 0.35  & 5024 $\pm$  87 & 2.52 $\pm$ 0.14 & -0.720 $\pm$ 0.05    \\ 
HD 145417\tablenotemark{a}  &    8.68 $\pm$ 0.37  & 4900           & 4.60            & -1.349               \\ 
HD 151559                     &   16.67 $\pm$ 0.37  & 5167 $\pm$  59 & 2.54 $\pm$ 0.13 & -0.769 $\pm$ 0.05    \\ 
HD 17072                      &   62.43 $\pm$ 0.36  & 5398 $\pm$  58 & 2.37 $\pm$ 0.14 & -1.128 $\pm$ 0.05    \\ 
HD 184266                     & -348.82 $\pm$ 0.47  & 5908 $\pm$  80 & 2.20 $\pm$ 0.18 & -1.463 $\pm$ 0.06    \\ 
HD 190287                     &  143.74 $\pm$ 0.33  & 5108 $\pm$  47 & 2.78 $\pm$ 0.12 & -1.429 $\pm$ 0.04    \\ 
HD 199289                     &   -5.95 $\pm$ 0.34  & 5842 $\pm$  54 & 4.27 $\pm$ 0.13 & -1.090 $\pm$ 0.05    \\ 
HD 211998                     &   32.10 $\pm$ 0.34  & 5233 $\pm$  46 & 3.25 $\pm$ 0.12 & -1.519 $\pm$ 0.04    \\ 
HD 219617                     &   13.28 $\pm$ 0.35  & 5981 $\pm$  61 & 4.40 $\pm$ 0.15 & -1.427 $\pm$ 0.05    \\ 
HD 221580                     &  -17.35 $\pm$ 0.36  & 5245 $\pm$  60 & 2.30 $\pm$ 0.15 & -1.215 $\pm$ 0.05    \\ 
HD 222434                     &   16.88 $\pm$ 0.36  & 4517 $\pm$  53 & 1.21 $\pm$ 0.13 & -1.770 $\pm$ 0.05    \\ 
HD 222925                     &  -38.64 $\pm$ 0.36  & 5710 $\pm$  60 & 2.32 $\pm$ 0.14 & -1.371 $\pm$ 0.05    \\ 
HD 23798                      &   88.88 $\pm$ 0.34  & 4396 $\pm$  60 & 0.64 $\pm$ 0.15 & -2.220 $\pm$ 0.05    \\ 
HD 26169                      &  -34.42 $\pm$ 0.35  & 5010 $\pm$  68 & 2.06 $\pm$ 0.16 & -2.396 $\pm$ 0.06    \\ 
HD 83212                      &  109.21 $\pm$ 0.32  & 4543 $\pm$  57 & 1.17 $\pm$ 0.13 & -1.504 $\pm$ 0.05    \\ 
HD 9051                       &  -71.99 $\pm$ 0.38  & 4884 $\pm$  49 & 2.08 $\pm$ 0.12 & -1.652 $\pm$ 0.04    \\ 
\enddata
\tablenotetext{a}{Reference star (see Section~\ref{sec:abundances} and Table~\ref{refstars}).}
\end{deluxetable*}

\subsection{MIKE Optical Spectra}

Optical spectroscopic observations of our sample stars were carried out using the MIKE spectrograph 
on the 6.5\,m Magellan/Clay Telescope at Las Campanas Observatory. We employed the standard setup 
with the narrowest slit (width 0.35 arcsec), leading to spectral resolution of $R\simeq83\,000$ 
and $65\,000$ in the blue and red CCDs, respectively. The full wavelength range covered by the 
standard setup of the MIKE spectrograph is 320-1000 nm. The S/N of our 
spectra varies from star to star, but it is typically in the range from 150 to 200 per pixel 
at 6000 {\AA}.

The MIKE spectra were reduced (trimmed, corrected for overscan, flat-fielded, and corrected for  
background/scattered light) with the CarnegiePython pipeline, which also provides 
a wavelength mapping on the data with the help of ThAr exposures taken throughout each observing 
night. These reduced spectra were further processed for continuum normalization, merging, and 
co-adding of multiple exposures (when needed) using common IRAF\footnote{IRAF is the Image 
Reduction and Analysis Facility, a general-purpose software system for the reduction and analysis 
of astronomical data. IRAF is written and supported by National Optical Astronomy Observatories 
(NOAO) in Tucson, Arizona.} tasks within the echelle package. 

The continuum normalization procedure can introduce errors in elemental abundance measurements. 
To estimate this effect, we tested different normalization parameters such as the order of the 
polynomials used to fit the upper envelopes of the data and the sigma-clipping values. On a 
representative spectrum from our dataset, we estimated changes in our EW measurements 
of up to 5\%, which translate to about 0.02 dex in [X/H]. These errors are smaller than our 
estimated total error of about 0.05 dex, which is dominated by the model uncertainties. Moreover, 
the changes in EW values were calculated relative to our preferred normalization method. If all 
EWs are measured consistently, i.e., on spectra that have been normalized using the same parameters, 
the errors are minimized because we employ strict differential analysis using stars with similar 
stellar parameters, for which the normalization procedure traces very similar continua. Even if the 
continuum errors were of order 0.02 dex, they would not be large enough to invalidate our main 
chemical tagging conclusions. As can be inferred from Figures~\ref{ONa} to \ref{Ba}, we would have 
had to under- or overestimate abundance ratios of some of our stars by up to 0.5 dex to make a 
number of our stars be consistent with the $\omega$ Centauri patterns.

\subsubsection{Radial Velocities (RVs)}\label{sec:rv}

RVs were computed by cross-correlation of our spectra with those of a set of 
five RV standard stars. Barycentric corrections were first computed using IRAF's task rvcor. 
The cross-correlation calculations were done employing IRAF's task fxcor. The standard stars were 
chosen from the RV catalogs by \cite{Nid02} and \cite{S13}. The standard stars are (their adopted 
RVs, in km s${}^{-1}$, are given in parenthesis, and are known to be stable within about 0.5 
km~s${}^{-1}$): HD~111721 ($21.4$), HD~128279 ($-75.8$), HD~193901 ($-171.5$), HD~134439 ($310.1$), 
and HD~134440 ($310.6$).

Cross-correlation was performed on the spectra on an order-by-order basis and for the blue and red 
CCDs separately. We employed 28 (15) orders in the blue (red) CCD, which cover the spectral range 
from 3560 to 4980 \AA\ (4850-6200 \AA). The order-to-order RV scatter of a typical cross-correlation 
calculation is less than about 0.2 km s${}^{-1}$. The RVs from the blue and red CCDs were found to be 
in excellent agreement, with a mean difference of $-0.14\pm0.36$ km s${}^{-1}$. Thus, we averaged the 
RVs from the blue and red CCDs to increase the RV precision.

Each stellar spectrum was cross-correlated with the five standards, resulting in five RV values. The final 
RV is the average of those five RVs, and the adopted error is the standard deviation from that mean. 
On average, this error is about 0.4 km s${}^{-1}$, which shows that our RV uncertainty is dominated 
by our choice of RV standards and, specifically, their adopted RV values.

\subsection{Infrared Spectra from the CRyogenic InfraRed Echelle Spectrograph (CRIRES)}\label{sec:crires}

High-resolution, high-S/N near-infrared spectra for the target stars were collected using the 
CRIRES \citep{K04} mounted on the VLT UT1 telescope. The observations were carried out in Service 
Mode (Program 090.B-0605(B), 29 hr, P.I. J.C.) from 2011 October through 2012 March.

The CRIRES setting employed a slit width of 0\farcs4 with the grating order \#52, yielding a spectral 
resolution of $R\simeq$~50\,000\,. This configuration allows us to observe the \hei\ 10830 {\AA} 
near the center of chip 3, which covers a range not affected by vignetting\footnote{Section 15 of CRIRES 
User Manual \url{http://www.eso.org/sci/facilities/paranal/instruments/crires/doc/VLT-MAN-ESO-14500-3486_V93.pdf}}. 
Under this setup, detector 3 has a wavelength reference of 10827 {\AA} (wavelength at the middle 
of the detector) and a wavelength coverage from 10800 to 10860 {\AA}.

Data reduction was performed using the CRIRES pipeline (Common Pipeline Library recipes, version 
2.4), running under the {\tt GASGANO} environment\footnote{Gasgano is an ESO tool designed to 
manage and handle in a systematic way the astronomical data observed and produced with VLT 
instruments. \url{http://www.eso.org/sci/software/gasgano/}}. Master dark subtraction, bad pixel 
map, flat-fielding, and corrections for nonlinearity effects were applied. Wavelength calibration 
for the spectra was applied using ThAr wavelamp frames. In order to improve the wavelength solution 
a high-resolution ISAAC spectrum\footnote{Available as Spectrocopic Standard at \url{http://www.eso.org/sci/facilities/paranal/decommissioned/isaac/tools/spectroscopic_standards.html}.} 
with telluric lines was used as reference and a conversion equation was determined from the 
difference between the nominal wavelength of the telluric lines and the obtained positions 
from the hot, fast rotating star HR~1996. On average, the telluric lines appeared redshifted 
by $\sim$~0.5\AA\ with respect to their nominal wavelength. 

Since the ThAr catalog lines used during the wavelength calibration of the spectra are in vacuum 
wavelengths, the \hei\ line does not appear at 10830 {\AA} (air wavelength), but 
instead it is found slightly redshifted, at 10833 {\AA}. In order to properly join the 
information from the Arcturus stellar lines atlas by \citet[in vacuum wavelengths]{H95} and the 
atmospheric water vapor lines listed by \citet[in air wavelengths]{B73}, the IAU standard equation 
of \cite{E53} was used to go from one to the other. The coefficients for the infrared regime 
were derived by \cite{P72}. Table~\ref{air} shows the air and vacuum wavelengths for the stellar 
and telluric lines expected to be found in the CRIRES spectra.

\begin{deluxetable}{lcc}
\tablecaption{Air and Vacuum Wavelengths for Stellar and Telluric Lines around the \hei\ line.\label{air}}
\tablecolumns{3}
\tablewidth{0pt}
\tablehead{\colhead{Line} & \colhead{$\lambda_{\rm air}$} & \colhead{$\lambda_{\rm vacuum}$} \\ 
\colhead{} & \colhead{({\AA})} & \colhead{({\AA})} } 
\startdata
\multicolumn{3}{c}{\textbf{Stellar Lines}}\\
\hline
\cri\   & 10801.36  &  10804.32 \\ 
\mgi\   & 10811.08  &  10814.05 \\ 
\cri\   & 10816.90  &  10819.86 \\ 
\fei\   & 10818.28  &  10821.26 \\ 
\si\   & 10821.18  &  10824.17 \\ 
\cri\   & 10821.66  &  10824.58 \\ 
\sii\   & 10827.09  &  10830.06 \\ 
\hei\   & 10830.34  &  10833.31 \\ 
\cai\   & 10833.38  &  10836.35 \\ 
\nai\   & 10834.87  &  10837.84 \\ 
\cai\   & 10838.97  &  10841.94 \\ 
\sii\   & 10843.85  &  10846.82 \\ 
\cai\   & 10846.79  &  10849.74 \\ 
\tii\   & 10847.63  &  10850.61 \\ 
\hline
\multicolumn{3}{c}{\textbf{Telluric Lines}}\\
\hline
       & 10815.95  &  10818.92 \\ 
       & 10832.10  &  10835.08 \\ 
       & 10833.98  &  10836.95 \\ 
       & 10838.03  &  10841.00 \\ 
       & 10840.81  &  10843.78 \\ 
       & 10843.19  &  10846.17
\enddata
\end{deluxetable}

Telluric lines were identified based on the spectrum of the hot, fast rotating star HR~1996. 
The wavelength-calibrated spectrum of this star is shown in Figure~\ref{hr1996} of the appendix, 
where the most prominent telluric lines (in the vacuum wavelength rest frame) are marked. 

Figures~\ref{crires1}, \ref{crires2}, \ref{crires3} and \ref{crires4} show all our CRIRES rest-frame 
spectra in the region of interest. For all the spectra, the normalization of the continuum 
was performed using the {\tt continuum} task of IRAF. Our science targets are indicated by the top 
blue spectrum on each panel, while the green spectrum, shown for comparison purposes, corresponds 
to the telluric standard star shifted in wavelength to match the telluric features in the science 
target. The left and right dashed lines marked the positions of the \sii\ and \hei\ lines, respectively.

\section{Stellar parameters and chemical abundances}
\label{sec:abundances}

\subsection{Stellar Parameters}\label{sec:sparam}

\begin{deluxetable*}{cccccc}[htb!]
 \tablecaption{Reference Stars for the Determination of Fundamental Atmospheric Parameters \label{refstars}}
 \tablecolumns{6}
\tablehead{\colhead{Reference Star} & \colhead{Group} & \colhead{$\teff$(K)} & \colhead{$\logg$} & \colhead{$\feh$} & \colhead{$\vt$} \\ \colhead{ } & \colhead{ } & \colhead{}  & \colhead{ } & \colhead{ } & \colhead{ } }
 \startdata
HD 113083 & dwarfs and subgiants   & 5800 & 4.35 & $-0.93$ & 1.20 \\
HD 145417 & cool dwarfs            & 4900 & 4.60 & $-1.30$ & 0.00 \\
HD 111721 & red giant branch stars & 4990 & 2.60 & $-1.38$ & 1.35 \\ 
 \enddata
\tablecomments{Conservatively, we expect these parameters to be uncertain at the level of 100 K in $\teff$, 0.1 in $\logg$, and 0.05 dex in $\feh$.} 
\end{deluxetable*}

The fundamental atmospheric parameters of our stars ($\teff,\logg,\feh$) were determined using the 
conditions of relative excitation and ionization equilibrium for iron lines (\fei\ and \feii). 
The EWs of as many as possible of the iron lines employed by \cite{R13} were measured 
by fitting Gaussian profiles to the observed spectral lines within IRAF's task {\tt splot}.

For each star, we started with a set of guess parameters (obtained from previous calculations 
found in the literature) and computed iron abundances using the curve-of-growth method on a 
line-by-line basis. Kurucz odfnew model atmospheres and the spectrum synthesis code MOOG were 
employed for these calculations. We then examined the correlations between relative iron abundance 
(with respect to a specific reference star, as explained below) and excitation potential as well 
as reduced EW ($REW=\log EW/\lambda$). We also examined the difference between mean 
iron abundance inferred from \fei\ and \feii\ lines separately. The initial set of parameters was 
then modified iteratively until the correlations disappeared and the \fei\ and \feii\ lines 
provided the same mean iron abundance. On average, 98 lines (80 \fei\ and 18 \feii) were used to 
derive the iron abundances. The minimum number of Fe transitions measured was 21 \fei\ and 
8 \feii\ for HD~215801.

When using the excitation/ionization method described above, it is customary to use the Sun as the  
standard star. However, our sample is heterogeneous regarding spectral type and it also is biased 
toward very low metallicity. Employing the Sun as reference is not ideal due to the propagation 
of systematic uncertainties, which are very difficult to control in highly heterogeneous samples 
like ours. Thus, instead we adopted three different reference stars that are representative of the 
three main groups of targets that we have available: dwarf and subgiant stars, cool dwarf stars, 
and RGB stars. The three reference stars and their adopted stellar parameters are 
listed in Table~\ref{refstars}.

For each of the reference stars we performed a relative analysis using the Sun as reference and 
adopting an absolute solar iron abundance of $A_\mathrm{Fe}^\odot=7.50$. Thus, while systematics 
still exist, they can be better controlled by investigating the parameters of the reference stars. 
The relative parameters within each group of stars are expected to be precise and internally 
consistent. In any case, the stars chosen as reference are all well-studied bright halo stars, 
and the parameters we derived using the Sun as reference are in reasonably good agreement with 
other previous and independent estimates. Conservatively, we expect these parameters to be uncertain 
at the level of 100 K in $\teff$, 0.1 in $\logg$, and 0.05 dex in $\feh$.

We note that the reference stars are on the metal-rich side of the stellar parameter distribution 
of their respective samples. Strictly speaking, it would be better to employ reference objects 
with stellar parameters that are centered on the distributions of each sample. However, good 
reference stars at low metallicities are significantly less common than metal-rich ones and, while 
not ideal, our procedure is certainly superior to one where the Sun, a very metal-rich star in 
this context, is employed as the only reference star.

Our derived stellar parameters are listed in Table~\ref{targets}. The errors were computed by 
propagating the errors in the slopes of the iron abundance versus excitation potential and reduced 
EW added in quadrature with the errors due to the uncertainty in the mean iron 
abundances from \fei\ and \feii\ lines measured separately. For $\feh$ we also added in quadrature 
the standard error of the line-to-line dispersion.

\begin{figure*}
\centering
 \includegraphics[width=0.75\linewidth]{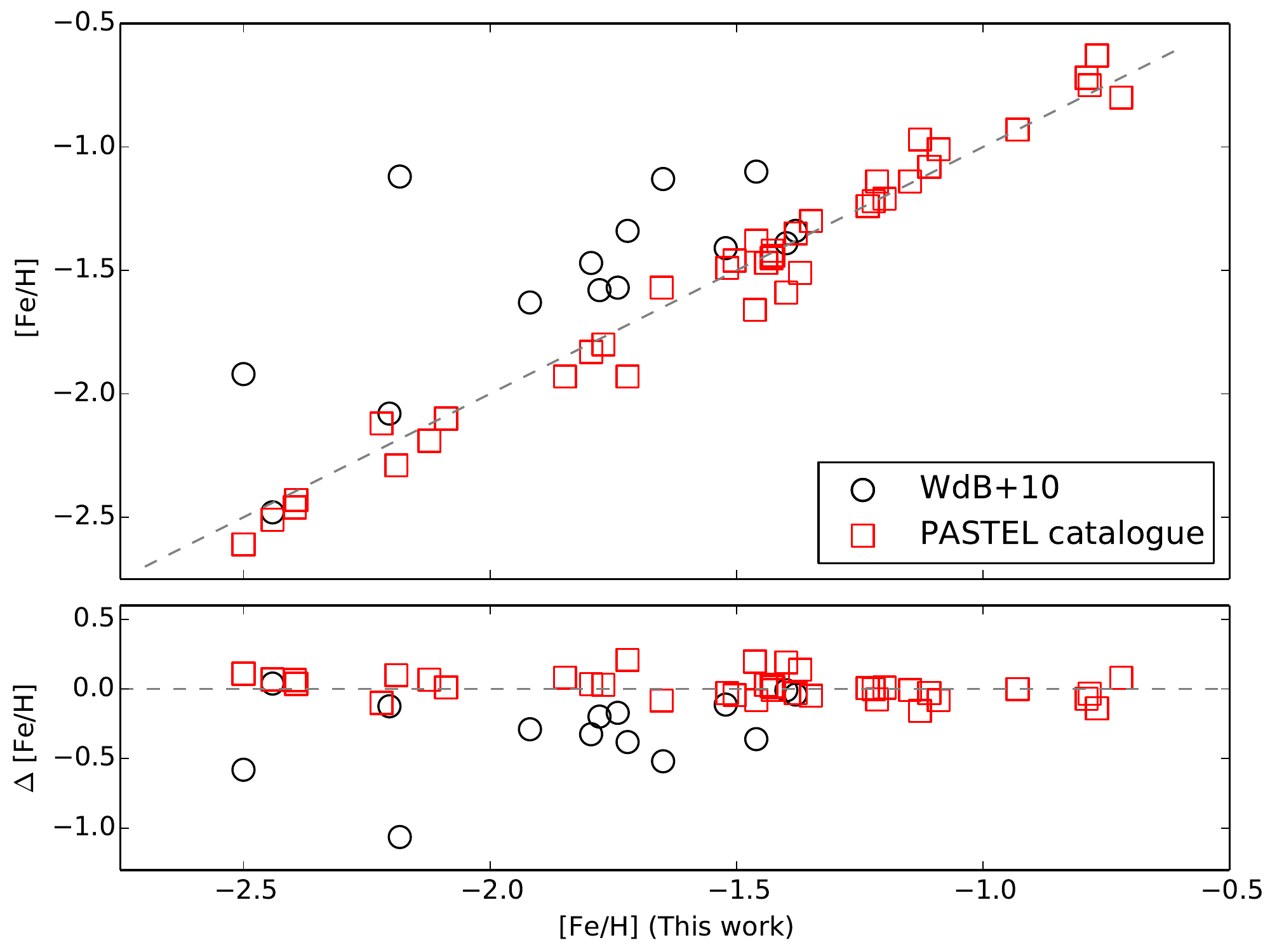}
 \caption{Comparison between our measured $\feh$ values and the derived values from \cite{W10} and the PASTEL catalog \citep{S10}. 
 {\it Top panel:} $\feh$ measurements in this work against the values measured by \cite{W10} for stars in the Kapteyn group, 
 and for stars in any of the three groups, corresponding to the average metallicity listed in the PASTEL catalog. 
 {\it Bottom panel:} the differences between the metallicity values measured in this work and the values found in the literature.  }
 \label{feh_comparison}
\end{figure*}

Figure~\ref{feh_comparison} shows our metallicity measurements for the stars that have also been 
listed by \citetalias{W10} and the PASTEL catalog \citep{S10}. The dashed line in the top panel corresponds 
to the 1:1 line, where practically all the stars with measurements from the PASTEL catalog lie. 
In the case of the stars with metallicity measurements in the study of \citetalias{W10}, the differences are 
higher ($\sim$0.3 dex). All the metallicities measured by those authors are higher than the ones 
measured in this work and in the PASTEL catalog.

\begin{deluxetable*}{lccccccr}
\tablecaption{List of Atomic Transitions and Equivalent Width Measurements for Our Target Stars.\label{ews_large}}
\tablecolumns{8}
\tablewidth{0pt}
\tablehead{
 & & & \multicolumn{5}{c}{Equivalent Width} \\
\cline{5-8} \\
 \colhead{Wavelength} & \colhead{Species} & \colhead{$\chi$} & \colhead{$\log{gf}$} & \colhead{CD-30 1121} & \colhead{CD-62 1346} & \colhead{HD 110621} & \colhead{(Cont.)}\\
 \colhead{({\AA})} & \colhead{} & \colhead{(eV)} & \colhead{} & \colhead{(m{\AA})}  & \colhead{(m{\AA})} & \colhead{(m{\AA})} & \colhead{}  
 } 
\startdata
7771.94  & 8.0    & 9.15  & 2.25          & 12.3       & 66.8       & 43.7      &  \\
7774.16  & 8.0    & 9.15  & 1.67          & 8.5        & 51.8       & 34.1      &  \\
7775.39  & 8.0    & 9.15  & 1.00          & 6.0        & 37.7       & 27.7      &  \\
4751.82  & 11.0   & 2.10  & -2.08         & \nan       & \nan       & \nan      &  \\
4982.82  & 11.0   & 2.10  & -1.00         & \nan       & \nan       & \nan      &  \\
5682.64  & 11.0   & 2.10  & -0.77         & 6.4        & 13.0       & 6.9       &  \\
5688.21  & 11.0   & 2.10  & -0.48         & 13.1       & 25.1       & 10.1      &  \\
6154.23  & 11.0   & 2.10  & -1.55         & \nan       & \nan       & \nan      &  \\
6160.75  & 11.0   & 2.10  & -1.25         & \nan       & 6.8        & \nan      &  \\
4730.04  & 12.0   & 4.34  & -2.39         & 8.6        & 24.3       & 7.2       &  \\
5711.09  & 12.0   & 4.34  & -1.73         & 27.8       & 41.9       & 19.7      &  \\
6318.72  & 12.0   & 5.10  & -1.95         & 5.1        & 11.3       & \nan      &  \\
 \enddata
\tablecomments{(This table is available in its entirety in machine-readable form.)}  
\end{deluxetable*}

The differences in metallicity between the measurements of \citetalias{W10} and the values derived in this 
work and the PASTEL catalog are propagated into the stellar parameters. For example, the stellar 
parameters of the star HD~181007 derived by \citetalias{W10} and the ones listed in the PASTEL catalog are 
in disagreement by more than 250 K for $\teff$ and 1.0 and 0.5 dex for $\logg$ and $\feh$, 
respectively. More worrying cases include HD~186478 and HD~181743. For the former, \citetalias{W10} report 
$\feh = -1.92$ while four studies included in PASTEL reveal a value closer to $-2.6$ dex. In the 
case of HD~181743 the $\feh$ and $\logg$ values given by \citetalias{W10} are 1.1 and 0.2 dex lower, 
respectively, than most previously reported measurements, while the $\teff$ is in good agreement 
with the values reported by the PASTEL catalog.

Moreover, when comparing our $\feh$ values with those from the PASTEL catalog, we find an average 
difference of 0.01$\pm$0.09 dex. Our iron abundances have a typical error of the order of 0.05 dex, whereas 
a heterogeneous catalog like PASTEL is expected to have $\feh$ errors in the range 0.05--0.10 dex. 
Thus, the predicted error in the average difference quoted above should be between 0.07 and 0.11 dex, 
in good agreement with the observed value of 0.09 dex. In stark contrast, when the \citetalias{W10} data are 
compared to the PASTEL catalog, the average difference in $\feh$ is 0.31$\pm$0.26 dex. The large 
dispersion seen in this comparison, and the small one seen in our case, strongly suggest that the 
$\feh$ values from \citetalias{W10} are unreliable.

\subsection{Elemental Abundances from the Optical Spectra}\label{sec:abund}

\begin{deluxetable*}{cccccccccccccc}
\tablecaption{Chemical Abundances for Our Sample Stars\label{abundances}}
\tablecolumns{14}
\tablewidth{0pt}
\tablehead{\colhead{Star} & \colhead{[Fe/H]} & \colhead{[O/Fe]} & \colhead{$\sigma$} & \colhead{[Na/Fe]} & \colhead{$\sigma$} & \colhead{[Mg/Fe]} & \colhead{$\sigma$} & \colhead{[Al/Fe]} & \colhead{$\sigma$} & \colhead{[Ca/Fe]} & \colhead{$\sigma$} &\colhead{[Ba/Fe]} & \colhead{$\sigma$} \\
 \colhead{} & \colhead{} & \colhead{} & \colhead{} & \colhead{} & \colhead{} & \colhead{} & \colhead{} & \colhead{} & \colhead{} & \colhead{} & \colhead{} & \colhead{} & \colhead{} } 
\startdata
\multicolumn{14}{c}{\textbf{Kapteyn Group Stars}}\\
\hline
 CD-30 1121  & -1.919 &  0.33*  & 0.07(3) & -0.15 & 0.06(2) & 0.23 & 0.06(3) & \nodata & \nodata & 0.25 & 0.08(9)  & -0.08 & 0.10(3) \\
 CD-62 1346  & -1.399 &  0.35   & 0.08(3) &  0.00 & 0.11(3) & 0.39 & 0.21(4) & \nodata & \nodata & 0.25 & 0.09(12) &  1.59 & 0.07(3) \\
   HD 110621 & -1.522 &  0.42   & 0.09(3) & -0.05 & 0.08(2) & 0.29 & 0.08(3) & \nodata & \nodata & 0.34 & 0.09(10) &  0.05 & 0.09(3) \\
   HD 111721 & -1.380 &  0.48   & \nodata &  0.04 & \nodata & 0.39 & \nodata & 0.26    & \nodata & 0.38 & \nodata  &  0.10 & \nodata \\
    HD 13979 & -2.441 & \nodata & \nodata & -0.21 & 0.09(1) & 0.43 & 0.10(3) & \nodata & \nodata & 0.29 & 0.11(7)  & -0.50 & 0.26(3) \\ 
   HD 181007 & -1.721 &  0.43*  & 0.07(3) &  0.15 & 0.12(6) & 0.35 & 0.09(3) & \nodata & \nodata & 0.33 & 0.07(13) &  0.30 & 0.05(3) \\
   HD 181743 & -1.795 &  0.57*  & 0.11(3) & -0.20 & 0.09(3) & 0.35 & 0.10(3) & \nodata & \nodata & 0.30 & 0.11(9)  & -0.16 & 0.09(3) \\
   HD 186478 & -2.500 &  0.72*  & 0.09(2) &  0.11 & 0.08(2) & 0.45 & 0.11(3) & \nodata & \nodata & 0.37 & 0.09(12) & -0.34 & 0.10(3) \\
   HD 188031 & -1.649 &  0.43*  & 0.06(3) & -0.09 & 0.08(1) & 0.18 & 0.07(2) & \nodata & \nodata & 0.36 & 0.08(7)  & -0.06 & 0.08(3) \\
   HD 193242 & -1.778 &  0.55*  & 0.07(3) &  0.01 & 0.09(3) & 0.37 & 0.08(3) & 0.13    & 0.07(1) & 0.35 & 0.08(14) &  0.02 & 0.09(3) \\
   HD 208069 & -1.741 &  0.38*  & 0.06(3) & -0.19 & 0.08(3) & 0.18 & 0.05(3) & \nodata & \nodata & 0.26 & 0.06(12) & -0.16 & 0.06(3) \\
    HD 21022 & -2.204 &  0.59*  & 0.06(3) & -0.03 & 0.05(2) & 0.42 & 0.10(3) & \nodata & \nodata & 0.31 & 0.07(10) & -0.08 & 0.09(3) \\
   HD 215601 & -1.460 &  0.47   & 0.04(3) &  0.10 & 0.11(6) & 0.37 & 0.04(5) & 0.08    & 0.08(2) & 0.24 & 0.06(15) & -0.03 & 0.10(3) \\
   HD 215801 & -2.183 &  0.45*  & 0.09(3) & -0.14 & 0.08(1) & 0.20 & 0.06(2) & \nodata & \nodata & 0.46 & 0.08(8)  & -0.38 & 0.10(3) \\
\hline
\multicolumn{14}{c}{\textbf{$\omega$~Cen Group Stars}}\\
\hline
 BD+02 3375  & -2.190 & 0.76* & 0.10(3) &  0.12    & 0.08(1) & 0.28 & 0.07(2) & \nodata & \nodata & 0.37 & 0.08(7)  & -0.47 & 0.07(2) \\
 CD-61 0282  & -1.234 & 0.44  & 0.07(3) & -0.20    & 0.12(2) & 0.26 & 0.09(3) & -0.02   & 0.08(2) & 0.23 & 0.08(11) & -0.04 & 0.08(3) \\
   HD 113083 & -0.930 & 0.31  & \nodata &  0.01    & \nodata & 0.16 & \nodata & -0.02   & \nodata & 0.18 & \nodata  &  0.07 & \nodata \\
   HD 121004 & -0.784 & 0.54  & 0.04(3) &  0.15    & 0.07(6) & 0.33 & 0.06(2) &  0.32   & 0.04(4) & 0.29 & 0.06(15) &  0.03 & 0.04(3) \\
   HD 140283 & -2.393 & 0.55* & 0.08(3) &  0.06    & 0.06(2) & 0.19 & 0.07(1) & \nodata & \nodata & 0.20 & 0.07(6)  & -0.92 & 0.05(2) \\
   HD 148816 & -0.790 & 0.45  & 0.04(3) &  0.19    & 0.06(6) & 0.33 & 0.07(4) &  0.22   & 0.06(5) & 0.23 & 0.05(14) & -0.09 & 0.06(3) \\
   HD 193901 & -1.109 & 0.40  & 0.05(3) & -0.20    & 0.06(4) & 0.09 & 0.08(3) &  0.07   & 0.06(2) & 0.19 & 0.06(10) & -0.03 & 0.06(3) \\
   HD 194598 & -1.148 & 0.36  & 0.04(3) & -0.03    & 0.06(3) & 0.14 & 0.09(4) & \nodata & \nodata & 0.21 & 0.06(13) & -0.05 & 0.04(3) \\
     HD 3567 & -1.200 & 0.29  & 0.04(3) & -0.13    & 0.10(4) & 0.12 & 0.06(2) & \nodata & \nodata & 0.28 & 0.06(12) &  0.03 & 0.05(3) \\
    HD 84937 & -2.089 & 0.53* & 0.08(3) &  \nodata & \nodata & 0.20 & 0.09(1) & \nodata & \nodata & 0.33 & 0.08(9)  & -0.20 & 0.18(2) \\
\hline
\multicolumn{14}{c}{\textbf{Field Stars}}\\
\hline
   HD 102200 & -1.222 &  0.30   & 0.06(3) & -0.06 & 0.05(2) & 0.24 & 0.09(5) & \nodata & \nodata & 0.24 & 0.08(12) & -0.06 & 0.06(3) \\
   HD 116064 & -1.848 &  0.50*  & 0.07(3) &  0.13 & 0.13(3) & 0.30 & 0.07(2) & \nodata & \nodata & 0.55 & 0.08(6)  & -0.23 & 0.07(3) \\
   HD 128279 & -2.123 &  0.41*  & 0.13(3) & -0.19 & 0.11(2) & 0.23 & 0.09(4) & \nodata & \nodata & 0.30 & 0.10(10) & -0.61 & 0.11(3) \\
   HD 134439 & -1.440 & \nodata & \nodata & -0.39 & 0.07(4) & 0.02 & 0.07(2) & \nodata & \nodata & 0.15 & 0.07(15) & -0.25 & 0.09(3) \\
   HD 134440 & -1.426 & \nodata & \nodata & -0.38 & 0.08(3) & 0.03 & 0.09(2) & \nodata & \nodata & 0.16 & 0.07(15) & -0.23 & 0.08(3) \\
   HD 142948 & -0.720 &  0.38   & 0.05(3) &  0.25 & 0.09(6) & 0.41 & 0.08(4) & 0.38    & 0.06(5) & 0.22 & 0.06(15) & -0.13 & 0.05(3) \\
   HD 145417 & -1.349 &  0.63   & \nodata &  0.02 & \nodata & 0.33 & \nodata & 0.23    & \nodata & 0.37 & \nodata  &  0.05 & \nodata \\
   HD 151559 & -0.769 &  0.17   & 0.05(3) &  0.18 & 0.07(6) & 0.21 & 0.05(4) & 0.20    & 0.09(5) & 0.17 & 0.06(15) &  0.07 & 0.05(3) \\
    HD 17072 & -1.128 &  0.50   & 0.05(3) &  0.07 & 0.07(6) & 0.38 & 0.05(4) & 0.29    & 0.09(4) & 0.25 & 0.07(13) &  0.31 & 0.09(3) \\
   HD 184266 & -1.463 &  0.51   & 0.11(3) & -0.01 & 0.10(2) & 0.34 & 0.12(2) & \nodata & \nodata & 0.28 & 0.08(11) &  0.24 & 0.32(3) \\
   HD 190287 & -1.429 &  0.53   & 0.04(3) &  0.03 & 0.09(5) & 0.38 & 0.06(5) & 0.22    & 0.10(2) & 0.41 & 0.07(15) &  0.09 & 0.05(3) \\
   HD 199289 & -1.090 &  0.52   & 0.05(3) &  0.11 & 0.15(5) & 0.33 & 0.06(4) & 0.17    & 0.06(3) & 0.25 & 0.06(13) & -0.09 & 0.06(3) \\
   HD 211998 & -1.519 &  0.52   & 0.04(3) &  0.00 & 0.09(4) & 0.41 & 0.09(4) & 0.04    & 0.06(1) & 0.41 & 0.06(15) & -0.03 & 0.06(3) \\
   HD 219617 & -1.427 &  0.41   & 0.06(3) & -0.11 & 0.06(2) & 0.18 & 0.08(3) & \nodata & \nodata & 0.23 & 0.06(9)  & -0.15 & 0.06(3) \\
   HD 221580 & -1.215 &  0.29   & 0.06(3) &  0.02 & 0.10(5) & 0.29 & 0.11(5) & 0.05    & 0.07(1) & 0.29 & 0.08(15) &  0.13 & 0.13(3) \\
   HD 222434 & -1.770 &  0.42*  & 0.05(3) &  0.02 & 0.09(5) & 0.40 & 0.05(3) & 0.08    & 0.24(2) & 0.29 & 0.06(14) &  0.05 & 0.07(3) \\
   HD 222925 & -1.371 &  0.42   & 0.07(3) &  0.00 & 0.08(3) & 0.24 & 0.10(3) & \nodata & \nodata & 0.23 & 0.06(11) &  0.85 & 0.20(3) \\
    HD 23798 & -2.220 &  0.67*  & 0.10(3) &  0.16 & 0.11(3) & 0.51 & 0.07(4) & 0.11    & 0.07(1) & 0.35 & 0.08(12) & -0.31 & 0.06(3) \\
    HD 26169 & -2.396 &  0.68*  & 0.10(2) &  0.38 & 0.12(2) & 0.47 & 0.08(2) & \nodata & \nodata & 0.37 & 0.07(10) & -0.35 & 0.11(3) \\
    HD 83212 & -1.504 &  0.36   & 0.05(3) & -0.01 & 0.07(6) & 0.42 & 0.06(4) & 0.14    & 0.08(4) & 0.24 & 0.06(14) &  0.11 & 0.08(3) \\
     HD 9051 & -1.652 &  0.49*  & 0.07(3) & -0.03 & 0.06(4) & 0.32 & 0.06(4) & \nodata & \nodata & 0.32 & 0.05(14) &  0.23 & 0.05(3) \\
\enddata
\tablecomments{The number of lines used to derive the abundances of each element is listed in parenthesis. Starred [O/Fe] abundances have non-LTE corrections applied.} 
\end{deluxetable*}

We searched for as many as possible of the spectral lines listed in Table~1 of \cite{R11} and 
Table~4 of \cite{R14} in order to measure chemical abundances of elements other than iron. 
Since that list was constructed for stars of solar metallicity, many of those lines are too weak 
in the spectra of our target stars. Thus, the line list was expanded with spectral lines listed 
in the studies by \cite{N02, R03, N09}, and \cite{B13}. The first of these references provides a 
thorough and reliable list of lines for very metal-poor stars in general and the other three 
references, while made for studies of somewhat metal-rich stellar populations, include spectral 
lines that are also useful in the analysis of stars with effective temperatures and surface 
gravities significantly different from the solar values. On average, the number of transitions 
employed to derived the chemical abundances was three lines of each of the elements considered 
(\nai, \mgi, \ali, \baii, and \oi). For \cai\ the number of lines employed was, on average, 
ten per star. The individual EW measurements are listed in Table~\ref{ews_large}.

With the stellar parameters listed in Table~\ref{targets}, we used the abfind driver of MOOG and 
Kurucz odfnew model atmospheres to derive elemental abundances for our sample stars. As for the 
stellar parameters, the absolute abundances of the reference stars were first computed using the 
Sun as reference. Then, we measured only relative abundances for the rest of our targets, using 
the appropriate reference star for each group. Table~\ref{abundances} presents the metallicity 
and the derived chemical abundances [X/Fe] for our target stars. The number of lines on which 
the quoted errors are based is indicated in parenthesis.

The sodium lines in the spectrum of HD~84937 were all found to have very low local S/N and in most 
cases the line was not even detected. Thus, no Na abundance is given for this star.

Abundances of Al are given only for a limited number of stars due to the difficulty in measuring 
the EWs of the very weak Al lines available in our optical spectra. In fact, in most 
of the cases where no Al abundance is reported in Table~\ref{abundances}, the Al lines were not 
even detected.

\begin{figure*}[htb!]
 \centering
 \includegraphics[width=0.75\linewidth]{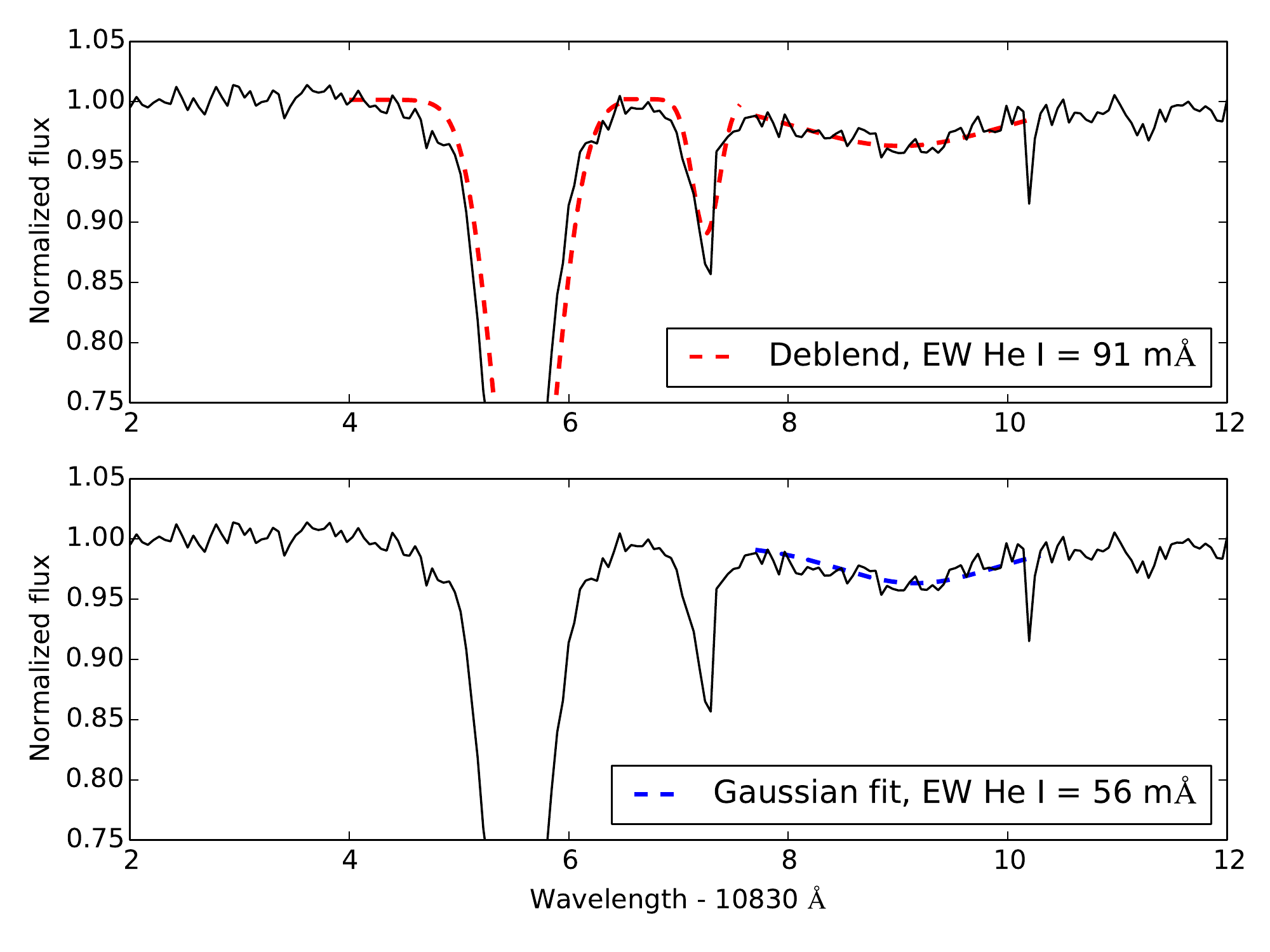}
 \caption{Fit for the \hei\ line in the infrared spectrum of CD-62 1346. The top panel shows the continuum-normalized CRIRES 
 spectrum in which the deblending option was used to fit simultaneously the widths of the three prominent absorption lines. 
 The bottom panel shows the same spectrum and the direct Gaussian fit for the \hei\ line.}
 \label{HeFit}
\end{figure*}

Given the importance of the oxygen abundances for this work, and the fact that they were derived 
using the 777\,nm O\,\textsc{i} triplet, which is known to be severely affected by departures from 
the LTE approximation, we made every effort to obtain as reliable as possible [O/Fe] values. 
We employed the grid of non-LTE corrections by \cite{R07} to derive line-by-line non-LTE oxygen 
abundances of all stars more iron-rich than $\feh=-1.6$. This grid has a lower limit in metallicity 
of $\feh=-1.4$, but it has been shown to be reliable in extrapolations down to $\feh=-1.6$ 
\citep[see, for example,][]{R11, R13}. The non-LTE corrected oxygen abundances of our three 
reference stars were computed differentially with respect to the Sun. Then, relative oxygen 
abundances were calculated for the other $\feh>-1.6$ stars differentially, each with respect 
to its corresponding reference star. Note that all our three reference stars have $\feh>-1.4$.

A significant fraction of our stars are more iron-poor than $\feh=-1.6$. For these objects, 
we employed LTE abundances to calculate oxygen abundances relative to their corresponding 
reference stars. However, for the reference stars, we adopted the non-LTE oxygen abundances 
derived as described above. This mixing of LTE and non-LTE oxygen abundances will certainly 
introduce a systematic error, but it will be somewhat mitigated by the fact that the calculations 
are done differentially for each star relative to another object of comparable atmospheric 
parameters. These ``pseudo'' non-LTE corrected oxygen abundances were adopted for the $\feh<-1.6$ stars.

To test the reliability of our adopted oxygen abundances, we computed non-LTE corrections for all 
our stars using the grid by \cite{F09}, which includes the full $\feh$ range covered by our stars. 
Nevertheless, we also had to extrapolate this grid in order to apply it to our data. This is 
because nearly all the LTE abundances we derived are higher than the upper limit given in that grid. 
In any case, we noticed that the [O/Fe] versus $\feh$ relation that results using the grid by 
\citeauthor{F09} is similar in shape to that obtained using our adopted oxygen abundances, even at 
low metallicity. In fact, the most notable difference is a roughly constant offset of about 
+0.15\,dex in [O/Fe] due to the fact that the oxygen abundances calculated using the grid by 
\citeauthor{F09} are systematically higher.

In a number of reliable elemental abundance studies, some of which employ oxygen features other 
than the 777\,nm triplet \citep[e.g.,][]{A01, S01, M02, M06}, the [O/Fe] abundance ratios of 
very metal-poor stars are seen to lie in a plateau at a level of about +0.5, similar to what we 
obtained with our adopted oxygen abundances. The [O/Fe] values inferred using the grid of 
\citeauthor{F09}, on the other hand, would place the [O/Fe] plateau at a significantly higher 
level ($\simeq+0.65$). For this reason, we prefer to adopt the oxygen abundances as described 
before, and not those computed with the \citeauthor{F09} grid. The comparison, however, shows 
that in a relative sense, the two methods provide similar results. This means that stars that are 
too oxygen-poor or too oxygen-rich relative to the bulk of halo stars could be equally identified 
using either set of [O/Fe] results.

\begin{deluxetable*}{lrrrc}[htb!]
\tablecolumns{5}
\tablewidth{0pt}
\tablecaption{EWs for the \hei\ 10830 {\AA} for CRIRES Spectra.\label{table_he}}
\tablehead{\colhead{Star} & \colhead{S/N}\tablenotemark{(a)} & \colhead{EW${}_1$} & \colhead{EW${}_2$} & \colhead{Notes} \\ 
\colhead{} & \colhead{} & \colhead{(m{\AA})} & \colhead{(m{\AA})} & \colhead{}  } 
\startdata
\multicolumn{5}{c}{\textbf{Kapteyn Group Stars}}\\
\hline
CD-30 1121 & 129 & 71.4$\pm$3.0      & 59.4$\pm$2.9 & \nan                                        \\
CD-62 1346 & 114 & 91.2$\pm$4.4      & 50.0$\pm$3.8 & \nan                                        \\
HD 110621  & 130 & 25.9$\pm$1.8      & \nan         & Blend, the line cannot be measured manually \\
HD 111721  & 142 & 48.8$\pm$2.5      & 40.7$\pm$2.4 & \nan                                        \\
HD 13979   & 161 & \nan              & \nan         & Blend with a telluric line                  \\
HD 181007  & 93  & 38.2$\pm$3.8      & 43.2$\pm$7.6 & Blend with a telluric line                  \\
HD 181743  & 84  & \nan              & \nan         & Blend with a telluric line                  \\
HD 186478  & 79  & $\leq$3.8$\pm$1.5 & \nan         & Upper limit                                 \\
HD 188031  & 71  & 18.5$\pm$3.0      & 10.9$\pm$2.3 & \nan                                        \\
HD 193242  & 83  & 41.4$\pm$4.4      & 49.0$\pm$4.6 & \nan                                        \\
HD 208069  & 125 & 62.7$\pm$3.5      & 50.1$\pm$3.1 & \nan                                        \\
HD 21022   & 163 & 12.3$\pm$1.4      & \nan         & Blend, the line cannot be measured manually \\
HD 215601  & 112 & \nan              & \nan         & Not detected                                \\
HD 215801  & 148 & 30.9$\pm$2.5      & 10.7$\pm$2.0 & \nan                                        \\
\hline
 \multicolumn{5}{c}{\textbf{$\omega$~Cen Group Stars}}\\
 \hline
BD+02 3375 & 81  & 23.3$\pm$3.2        & 23.1$\pm$2.9       & \nan          \\
CD-61 0282 & 140 & 21.6$\pm$1.7        & 21.8$\pm$1.9       & \nan          \\
HD 113083  & 137 & \nan                & \nan               & Not detected  \\
HD 121004  & 149 & 40.6$\pm$2.4        & 13.3$\pm$1.8       & \nan          \\
HD 140283  & 162 & $\leq$6.2 $\pm$1.3  & $\leq$10.0$\pm$1.9 & Upper limit   \\
HD 148816  & 171 & 35.1$\pm$2.5        & 23.1$\pm$2.3       & \nan          \\
HD 193901  & 123 & 31.9$\pm$2.7        & 31.5$\pm$2.8       & \nan          \\
HD 194598  & 104 & 31.1$\pm$3.5        & 34.4$\pm$3.7       & \nan          \\
HD 3567    & 118 & 36.9$\pm$3.5        & 31.3$\pm$2.8       & \nan          \\
HD 84937   & 127 & 27.6$\pm$2.8        & 25.4$\pm$2.9       & \nan          \\
\hline                                                        
 \multicolumn{5}{c}{\textbf{Field Stars}}\\
 \hline
HD 102200  & 129 & 25.9$\pm$2.6        & 26.3$\pm$2.9  & \nan                                        \\
HD 116064  & 125 & 11.8$\pm$1.8        & \nan          & Blend, the line cannot be measured manually \\
HD 128279  & 136 & 44.9$\pm$2.7        & 51.8$\pm$3.2  & \nan                                        \\
HD 134439  & 147 & 52.9$\pm$1.4        & \nan          & Blend, the line cannot be measured manually \\
HD 134440  & 147 & 34.7$\pm$1.6        & \nan          & Blend, the line cannot be measured manually \\
HD 142948  & 118 & 18.2$\pm$2.4        & 17.0$\pm$2.3  & \nan                                        \\
HD 145417  & 176 & 72.6$\pm$2.1        & 53.7$\pm$1.9  & \nan                                        \\
HD 151559  & 139 & 60.6$\pm$3.6        & 46.1$\pm$2.9  & \nan                                        \\
HD 17072   & 294 & 73.8$\pm$1.7        & \nan          & Blend, the line cannot be measured manually \\
HD 184266  & 99  & 442.8$\pm$6.1       & 302.5$\pm$5.3 & \nan                                        \\
HD 190287  & 136 & 44.5$\pm$2.8        & 49.8$\pm$2.8  & \nan                                        \\
HD 199289  & 116 & 10.7$\pm$1.9        & 14.5$\pm$2.2  & \nan                                        \\
HD 211998  & 521 & \nan                & \nan          & Blend with a telluric line                  \\
HD 219617  & 91  & \nan                & \nan          & Not detected                                \\
HD 221580  & 142 & 51.6$\pm$3.2        & 37.6$\pm$2.8  & \nan                                        \\
HD 222434  & 110 & $\leq$1.3$\pm$2.2   & \nan          & Upper limit                                 \\
HD 222925  & 110 & 231.4$\pm$5.1       & 215.2$\pm$4.5 & \nan                                        \\
HD 23798   & 161 & \nan                & \nan          & Emission?                                   \\
HD 26169   & 151 & 34.6$\pm$2.3        & 51.6$\pm$2.5  & \nan                                        \\
HD 83212   & 156 & \nan                & \nan          & Blend with a telluric line                  \\
HD 9051    & 129 & 31.7$\pm$2.0        & 46.2$\pm$3.7  & \nan                                        \\
\enddata
\tablenotetext{(a)}{S/N calculated as the median value of the S/N pixel by pixel.}
\end{deluxetable*}

The 777\,nm oxygen triplet lines were not detected in the spectra of HD~134439, HD~134440, and 
HD~13979. Therefore, no oxygen abundances are reported for these three stars.

Later in this paper, we use the elemental abundances derived from the optical spectra to perform a 
chemical tagging experiment, i.e., to attempt to associate our samples to known stellar populations 
for which previous chemical composition studies exist. Most of these previous works have adopted a 
pure LTE approach or used the oxygen forbidden lines, which are formed under LTE conditions, to 
derive [O/H]. Therefore, for our chemical tagging experiment, the impact of non-LTE effects on the 
abundances is not as important as being able to perform a fair comparison instead.

\subsection{\hei\ 10830 {\AA} line}
\label{sec:ew}

The EW of the \hei\ 10830 {\AA} line in our program stars was measured with the IRAF's task {\tt splot}. 
We use the deblend option when a telluric line appears contaminating the wings or the 
continuum around the line, which occurs in six of our program stars (four stars of the Kapteyn/$\omega$ 
Cen group sample and two of the field sample). For another star of the field comparison sample, 
HD~211998, a strong telluric line is at exactly the same position as the He line, preventing us from saying  
anything about it. 

We also measured, when possible, the EW of the line directly using the Gaussian fit of splot, 
as a validation test for our measured values. However, when telluric lines are near the \hei\ line, 
the position of the wings is not easy to identify and the line sometimes could be asymmetric, leading to  
a less accurate Gaussian fit. Figure~\ref{HeFit} shows the fit using both procedures for the \hei\ 
line of CD-621 346. From the spectrum, the wings of the \hei\ line are difficult to identify and the 
direct Gaussian fit gives a lower value than the deblending option, which takes into account the fact 
that part of the wings of one line and its neighbor can overlap.

The EW measurements derived using the deblending task, EW${}_1$, and the direct Gaussian fit, EW${}_2$, 
are listed in Table~\ref{table_he}, where we also list the S/N achieved by our 
CRIRES observations at the location of the \hei\ line. Errors in EW were calculated using the 
formula of \cite{C88}, which uses the FWHM of the line, the S/N of the spectrum, and the dispersion 
of the instrument (5.2 $\times 10^{-3}$ nm pix${}^{-1}$ for the setup used in our CRIRES observations). 
A minimum value of 10 m{\AA} as lower limit is used to separate stars that show detectable 
helium in their spectra from those that do not. A similar cut-off was found by \cite{D11} for 
the spectra of the stars where the helium line is not apparent.

In most of the spectra, there is a good agreement between the two EW values measured 
as mentioned before. The same behavior was found by \cite{Smith14}, comparing the EWs of the 
\hei\ line in giants of M13. Nonetheless, in some cases the difference is non-negligible and not explained 
by the listed errors, mainly because the line is in blends, and the Gaussian fit cannot take into account 
the part of the spectrum that is missed in the blended wing. Because of that, we adopt hereafter 
the EWs measured using the deblending option.

Although the \hei\ 10830{\AA} line has not been measured frequently in a systematic way, there 
are nine stars from our sample that have published EWs, which are listed in Table~\ref{ew_lit} 
together with our derived EW values from CRIRES spectra. A relatively good agreement is found for 
five stars; HD~194598 is the only star that has two measurements in the literature and both 
values agree with our derived value. However, there are some stars that have important 
differences, which might be due to an intrinsic variability of the line depending on the chromospheric 
activity. This is further discussed in Section~\ref{he_disc}.

\begin{deluxetable}{cccc}
\tablecaption{Comparison between \hei\ line EW measurements.\label{ew_lit}}
\tablecolumns{4}
\tablewidth{0pt}
\tablehead{\colhead{Star} & \colhead{EW (CRIRES)} & \colhead{EW (Lit.)} & \colhead{Reference} \\ 
\colhead{} & \colhead{(m{\AA})} & \colhead{(m{\AA})} & \colhead{} } 
\startdata
 \multicolumn{4}{c}{\textbf{Kapteyn group stars}}\\
 \hline
HD 111721 & 48.8 $\pm$2.5 & 48.4 & 1 \\
\hline
\multicolumn{4}{c}{\textbf{$\omega$~Cen-group stars}}\\
\hline 
HD 140283 &  6.2 $\pm$ 1.3 & 32.5 & 2 \\
HD 148816 & 31.1 $\pm$ 3.5 & 30.1 & 2 \\
HD 194598 & 31.1 $\pm$ 3.5 & 30.5, 26.7 & 2,3 f\\
HD 193901 & 31.9 $\pm$ 2.7 & 40.5 & 2 \\
HD 3567   & 36.9 $\pm$ 3.5 & 15.1 & 3 \\
HD 84937  & 27.6 $\pm$ 2.8 & 27.2 & 3 \\
 \hline
 \multicolumn{4}{c}{\textbf{Field stars}}\\
 \hline
HD 219617 & not detected & 25.3 & 2 \\
HD 83212  & not detected   & in emission & 1 \\
\enddata
\tablerefs{(1) \cite{D09}; (2) \cite{T11}; (3) \cite{S12}}
\end{deluxetable}

\section{Chemical abundance patterns}\label{sec:chemical}

\subsection{O--Na and Mg--Al}\label{ona_disc}

One of the main distinguishing chemical patterns of GC stars is an anticorrelated behavior of O and 
Na abundances, with a fraction of the stars showing O--Na levels similar to field stars, but 
another important fraction showing depleted O and enhanced Na \citep[][and references 
therein]{C09, JP10, G12}. The latter group of stars has been associated to a second stellar 
generation that formed out of material enriched by the hotter branches of hydrogen burning that 
took place in stars of the first generation \citep{Ventura01, G04}. Therefore, if the Galactic 
field contains a fraction of stars originating from GCs, one would expect to find some of 
these O-poor, Na-rich second-generation stars as well \citep[e.g.,][]{Altmann05, R12}. Applying this to the 
context of the present work, therefore, if the Kapteyn group were tidal debris from 
$\omega$~Centauri, then an unambiguous indication that this is the case would be the identification 
among the group's stars of some with this particular O--Na pattern.

\begin{figure*}[htb!]
\centering
\includegraphics[width=0.75\linewidth]{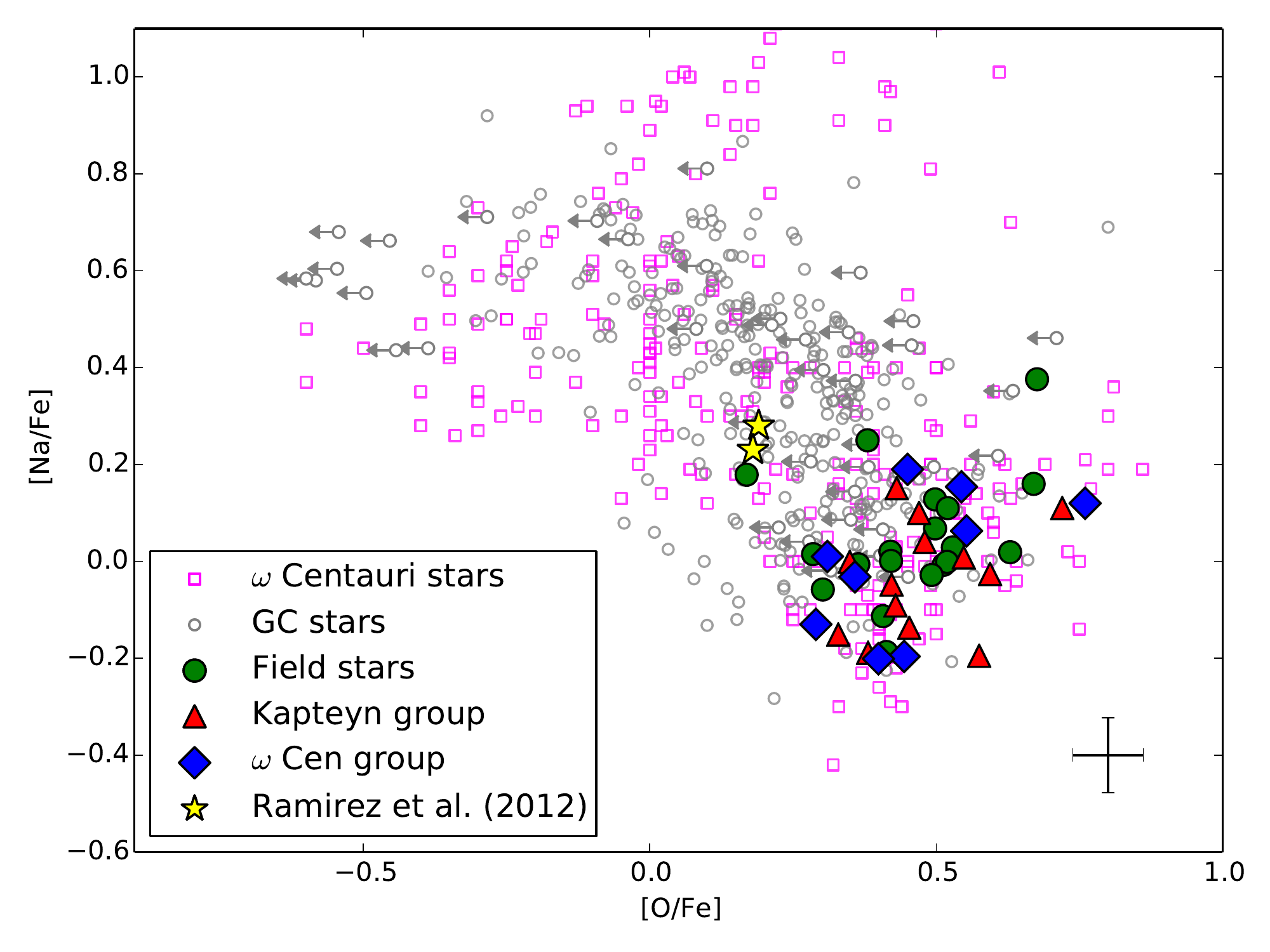}
\caption{Oxygen and sodium abundances for our target stars. Stars from the Kapteyn group and the 
$\omega$~Cen group are plotted as red triangles and blue diamonds, respectively, while the field 
stars are marked as green circles. For comparison purposes the O--Na abundances for GC stars from 
\cite{C09} are shown as open circles while the magenta squares are those derived by \cite{M11} 
for $\omega$~Centauri giants. Typical error bars for our measurements are shown at the bottom right 
corner.}
\label{ONa}
\end{figure*}

Figure~\ref{ONa} shows the O and Na abundances for our target stars from the Kapteyn group 
(red triangles), $\omega$~Cen group (blue diamonds) and field stars (green circles). 
The O and Na abundances of GC stars \citep[open gray circles from][]{C09} and giant stars from 
$\omega$~Centauri \citep[open magenta squares from][]{M11} are included. The two star symbols correspond 
to the two GC second-generation stars in the field found by \cite{R12}.

The oxygen abundances of the stars from both the Kapteyn and $\omega$~Cen groups are 
indistinguishable from those of the stars of the field sample, which are located at the base 
of the GC Na--O anticorrelation. HD~151559 appears as the only star from our whole sample that could be a 
second-generation star from a GC in the field. This star has the lowest O abundance and the third 
most enhanced Na abundance of our full sample.

$\omega$~Centauri is known to harbor at least three different populations, with mean metallicities 
at $\sim$~--1.7, --1.2 and --0.8 dex \citep{JP10}. These populations are consistent with the 
multiple MSs found through its color--magnitude diagram, namely the red, blue, and anomalous MS 
\citep[as defined by][]{B04, B10}. Spectroscopic analyses of RGB stars from these three different 
populations by \cite{M11} and \cite{G11} reveal that the metal-poor and metal-intermediate 
populations present a marked O--Na anticorrelation and at least two different generations of stars 
(primordial and intermediate, i.e., the first and second generations). It is important to note that 
this conclusion does not depend on the specific bins adopted to separate the three metallicity 
populations, thus discarding the possibility that the O--Na anticorrelation seen in both populations 
is due to the overlapping of the populations. Only the metal-rich population (at $\feh \sim -0.8$ dex) 
is practically all composed of second-generation stars and, contrary to the others, it actually 
shows a slight O--Na correlation \citep{G11}. In particular, \cite{G11} show that there is an 
extreme component of second-generation GC stars \citep[as defined by][]{C09} that is consistent with 
being O-poor, Na-rich, and also He-rich (up to $Y \simeq 0.387$ dex) compared to stars of an 
intermediate component that are also O-poor and Na-rich, but with almost the same He abundance as 
the primordial generation. This first generation is mostly indistinguishable from the halo field 
stars with the same metallicity. Similar to monometallic GCs, $\omega$~Centauri has a prominent 
fraction of second-generation stars \citep{JP10}. According to the criteria to separate the primordial, 
intermediate, and extreme components in GCs developed by \cite{C09} and the O and Na abundances 
reported by \cite{JP10}, the fractions of stars belonging to the primordial, intermediate, and extreme 
components in $\omega$~Centauri are 42\%, 44\%, and 14\%, respectively, in agreement with the measured 
fractions in monometallic GCs \citep{JP10}. Nonetheless, $\omega$~Centauri has a prominent fraction 
of extreme component stars in its metal-rich population, even reaching 81\% of that population, 
the greatest fraction found in any GC so far \citep{JP10}.

\begin{figure*}[htb!]
\centering
\includegraphics[width=0.75\linewidth]{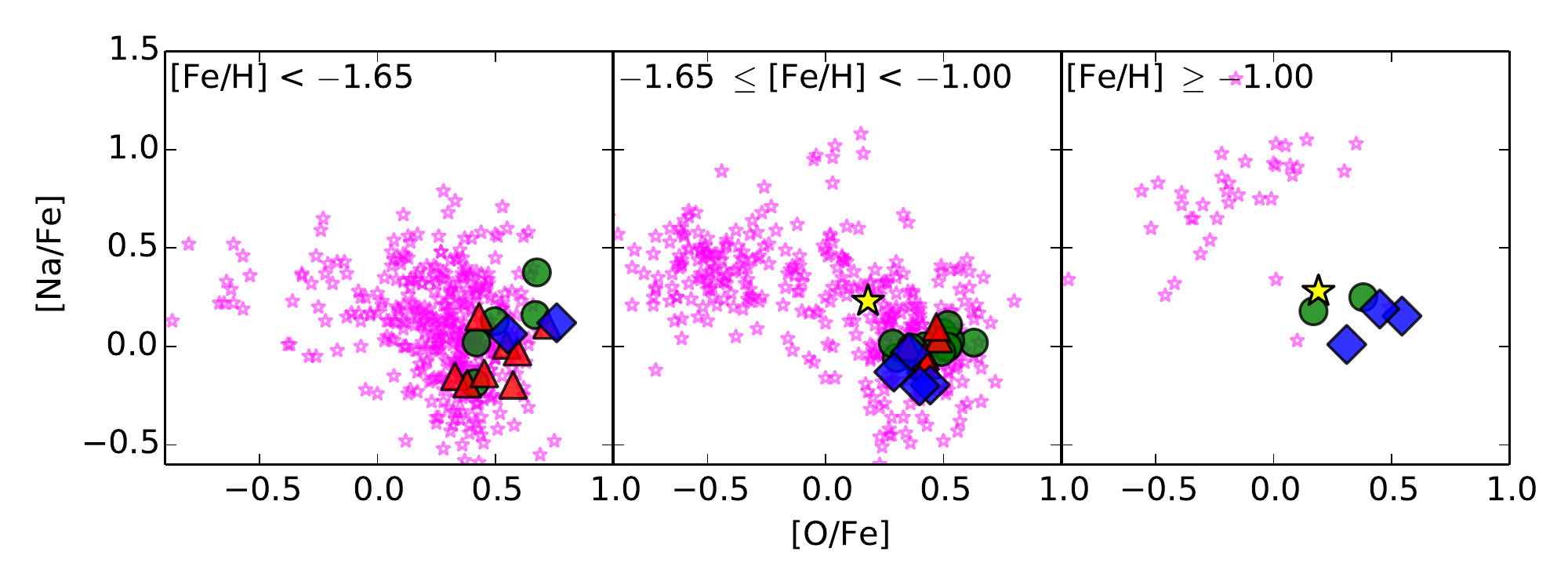}
\caption{O--Na anticorrelation for the three different metallicity groups present in 
$\omega$~Centauri (magenta stars). For each metallicity bin, the stars in our sample were 
plotted accordingly. The Kapteyn group has stars in the metal-poor and metal-intermediate groups. 
In the middle and right panels the two GC second-generation field stars found by \cite{R12} are also 
plotted, as solid yellow stars.}
\label{3panels}
\end{figure*}

Motivated by this, our target stars were divided into three different metallicity ranges, in order 
to better visualize whether second-generation stars are present among the Kapteyn or the 
$\omega$ Cen groups. Figure~\ref{3panels} shows the O--Na plane for the three different metallicity 
regimes, i.e., metal-poor, $\feh\ <-1.65$; metal-intermediate, $-1.65 \leq \feh\ < -1.00$; and 
metal-rich, $\feh\ \geq -1.00$. In each panel, $\omega$~Centauri's RGB stars from the catalogue of 
\cite{JP10} are also plotted, according to the different metallicity bins. 

From the metallicity distribution of the Kapteyn group stars, most of them (9 out 14 stars) are 
metal-poor, while five are in the metal-intermediate range and there are no stars in the metal-rich group. 
Figure~\ref{3panels} shows there is no evidence of an O--Na anticorrelation in the two groups. 
The same result is found for the stars in the $\omega$~Cen group, which are indistinguishable from 
the Galactic halo stars in each metallicity bin.

GC stars also follow a Mg--Al anticorrelation \citep{C09}, which is the result of high-temperature 
H-burning chains. In contrast, disk and halo stars of the MW are mostly confined to nearly 
solar Al and Mg abundances \citep{R03, R06}. Therefore, if stars in the Kapteyn and 
$\omega$~Cen groups showed that kind of behavior, it would be suggestive of a GC origin.

\begin{figure*}[htb!]
\centering
\includegraphics[width=0.75\linewidth]{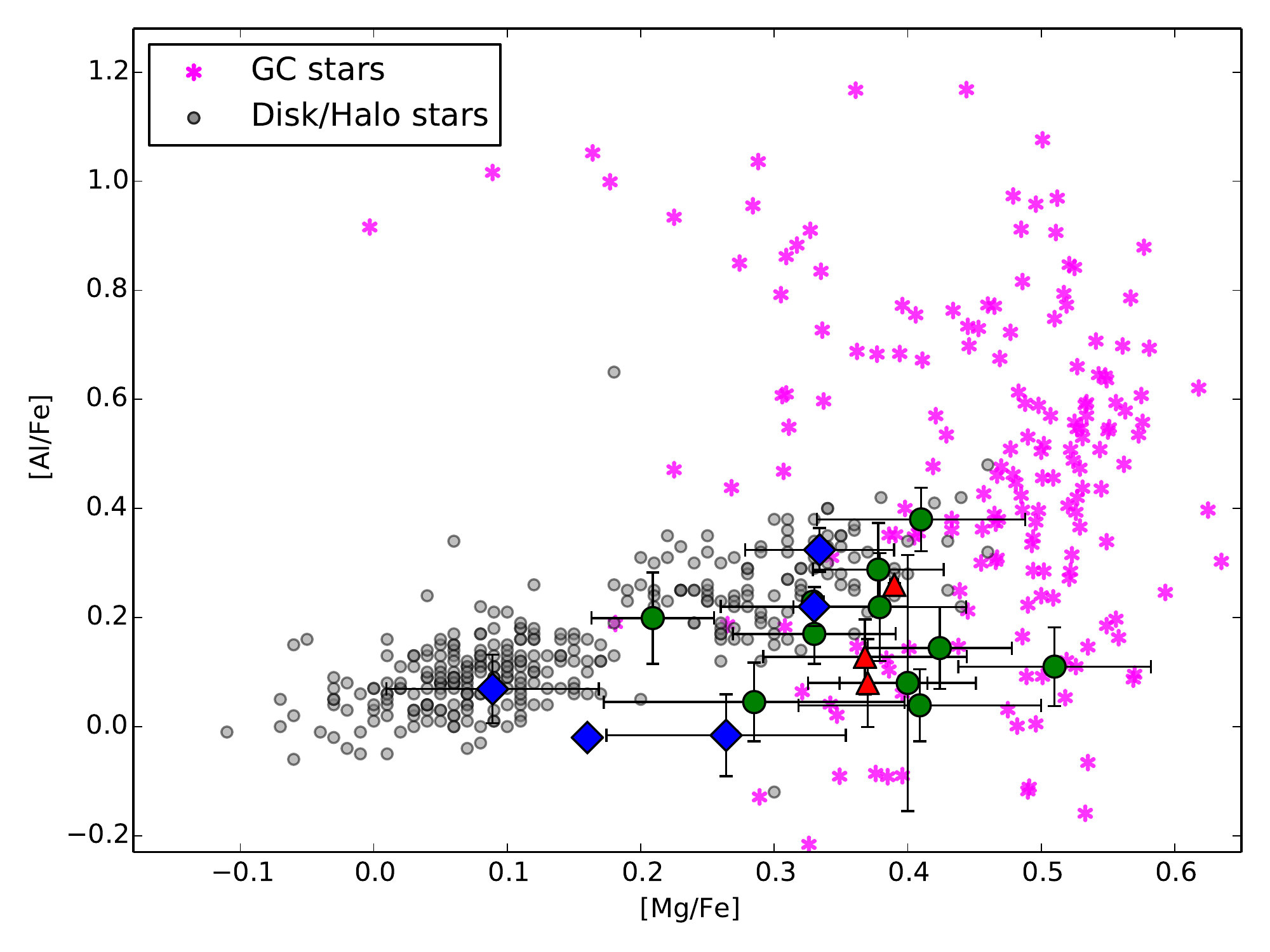}
\caption{Magnesium and aluminum abundances for our target stars (same symbols as in Figure~\ref{ONa}). 
Only three of the 14 stars observed from the Kapteyn group have aluminum abundances measured. 
Magenta starred symbols are GC stars from the catalog of \cite{C09}, which show a broad Mg--Al anticorrelation. 
Gray circles are galactic halo and disk stars from \cite{C00} and \cite{R03, R06}.}
\label{MgAl}
\end{figure*}

Aluminum lines are weak or undetectable for most of our target stars. Figure~\ref{MgAl} shows the 
stars for which we could measure an abundance for this species --only three from the Kapteyn group 
and five from the $\omega$~Cen one-- preventing us from tracing any possible Mg--Al pattern. All these stars, 
nevertheless, have similar abundances to those found for the field star samples. 

Could it be that our sample of Kapteyn group stars just so happens to contain only first-generation 
stars stripped from $\omega$~Centauri? As in other GCs, the second-generation stars in 
$\omega$~Centauri are more centrally concentrated than the first-generation \citep[see][and 
references therein]{B09}. Therefore, it can be expected that a larger fraction of the stripped stars 
from the cluster belong to the first generation. However, as we explain immediately below, if all 
our Kapteyn and $\omega$~Cen group stars originated in $\omega$~Centauri, it is unlikely that all 
of them could be at the same time from the first stellar generation of the cluster.

Considering that the fraction of second-generation GC stars in the Galactic halo is of the order of 
2-3\% \citep{Martell11, R12}, the models of \cite{V10} allow us to estimate that at most 50\% 
of the Galactic halo stars come from GCs. Nonetheless, if we consider that our stars are typical 
Galactic halo stars, from our observations, we have one out of 45 ($\simeq$ 2.2\%) stars with Na and O 
abundances consistent with GC second-generation stars, compatible with the 2-3\% fraction expected.

Based on the concentrations derived by \cite{G11}, the number of first- and 
second-generation stars from the different populations could be estimated out to 10 arcmin, 
corresponding to about a fourth of $\omega$~Centauri's tidal radius. According to Table 1 and 
Figure 9 of their paper, the mix of stars at 10 arcmin from the center should be composed of 35\%, 
60\%, and 5\% from the metal-poor, metal-intermediate, and metal-rich populations, respectively. 
Within the metal-poor and metal-intermediate populations, $\sim$30\% and 40\% of the stars should 
correspond to second-generation stars. Therefore, assuming the stripped regions of $\omega$~Centauri 
contained a similar mix of populations, if the Kapteyn and $\omega$~Cen groups were part of 
$\omega$~Centauri's tidal debris, they should contain approximately 30\% of second-generation 
stars: four and three stars in the Kapteyn and $\omega$~Cen groups, respectively. Although crude, 
this fraction of stars from the second-generation that could be stripped from $\omega$~Centauri 
is at odds with our derived chemical abundances, which are consistent with only first-generation stars.

\subsection{Helium} \label{he_disc}

\begin{figure*}[htb!]
\centering
\includegraphics[width=0.75\linewidth]{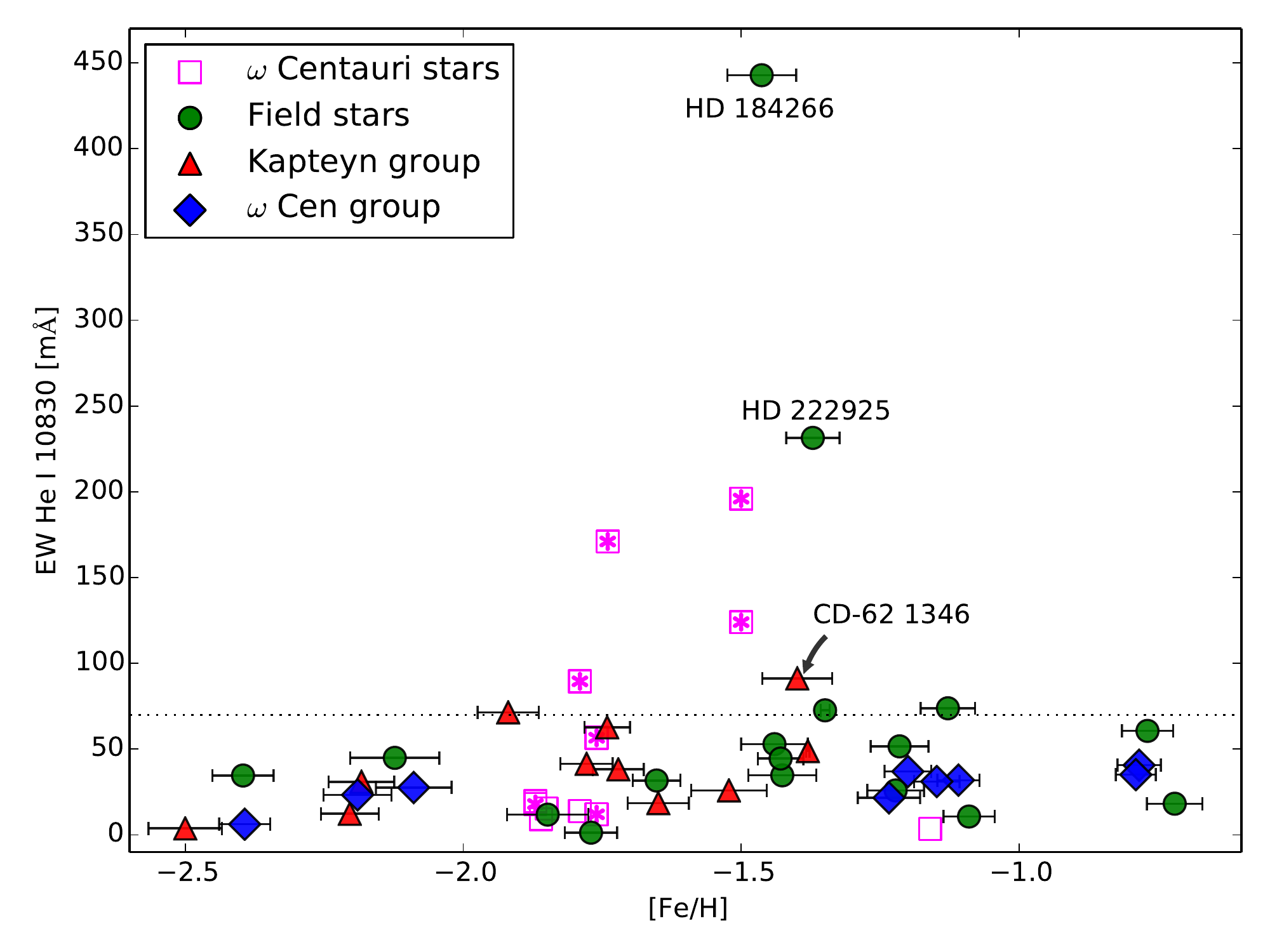}
\caption{EW for the \hei\ 10830{\AA} line vs. metallicity for our sample. The dotted line marks 
70 m{\AA}, an upper limit determined by \cite{S12} for non-active metal-poor field stars. The EW 
of \hei\ appears independent of the metallicity for the Kapteyn/$\omega$~Cen group stars. Magenta 
squares are for $\omega$~Centauri giant stars measured by \cite{D11}, while those with starred symbols 
correspond to enhanced Al and Na abundances.}
\label{ew_he}
\end{figure*}

Figure~\ref{ew_he} shows the EW of the \hei\ line for all the stars in our program as a function of 
metallicity. $\omega$~Centauri's red giants observed by \cite{D11} are also plotted as 
open magenta squares, with the starred symbols indicating those stars with enhanced Al and Na 
abundances. The dotted line at 70 m{\AA} marks the typical maximum value found in non-active 
metal-poor stars from the Galactic field \citep{T11, S12}. All the stars in the $\omega$~Cen group 
have EW below 70 m{\AA}, making the group indistinguishable from the sample of comparison field stars. 
Stars of the Kapteyn group also appear with typical EWs, except CD-62 1346 which has a somewhat 
large EW of 91.2 m{\AA}. There is no correlation between the EW of the He line and the metallicity, 
as seen for the metal-intermediate group of stars in $\omega$~Centauri \citep{D11}.

The $\omega$~Centauri giants with starred symbols in Figure~\ref{ew_he} are second-generation star 
candidates based on their Na and Al abundances \citep{D11}. These authors found a better correlation 
between the detection of the chromospheric \hei\ line and the enhanced Na--Al abundances than as a 
function of the metallicity. For stars in the Kapteyn and $\omega$~Cen groups, the measured \hei\ 
EWs are generally smaller than those of the secong-generation stars of \cite{D11} with seemingly 
enhanced helium. As discussed earlier, they also have O, Na, Mg, and Al abundances normal 
among field stars, which would be consistent with their \hei\ EWs, assuming the latter to be 
indicative of helium content. For red giants in Messier 13, a GC with no significant helium 
enhancement measured ($\Delta Y \sim$ 0.04), \cite{Smith14} found several \hei\ EW values lower 
than 40 m{\AA}, similar to the distribution found for the Kapteyn and $\omega$~Cen groups.

\begin{figure*}[htb!]
 \centering
\includegraphics[width=0.75\linewidth]{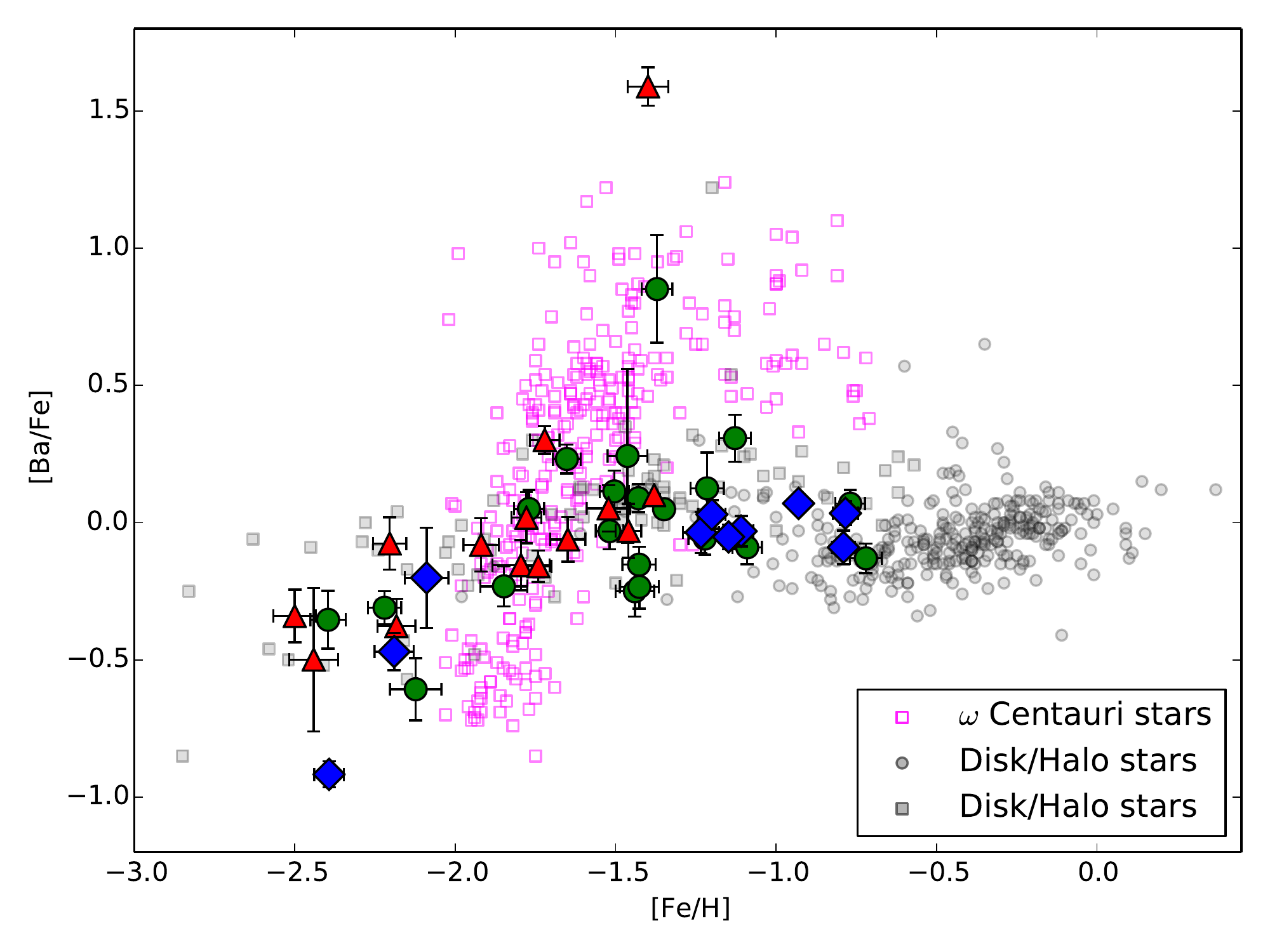}
\caption{[Ba/Fe] vs. [Fe/H] for our target stars (same symbols as in Figure~\ref{ONa}). Magenta 
open squares are $\omega$~Centauri stars from \cite{M11}, while gray circles and squares correspond 
to Milky Way stars from \cite{R03, R06} and \cite{I13}, respectively. CD-62 1346, from the Kapteyn 
group, and HD~222925, from our field control sample, are the stars with the highest Ba abundances.}
\label{Ba}
\end{figure*}

Among the control sample of field stars, HD~222925 and HD~184266 strongly deviate from all the 
stars in all of our samples. HD~184266 has an EW of the order of magnitude of those found in active 
cool stars \citep[see][]{S08}, but no previous studies reported a measurement for the helium line 
for this star. HD~222925 has a \hei\ EW about 5-6 times larger than the average value for the 
remaining stars in our field sample and, moreover, it has an enhanced [Ba/Fe] abundance 
(Fig.~\ref{Ba}) that makes it an intriguing target in its own right (see Section~\ref{peculiar}). 

\begin{figure*}[htb!]
\centering
\includegraphics[width=0.75\linewidth]{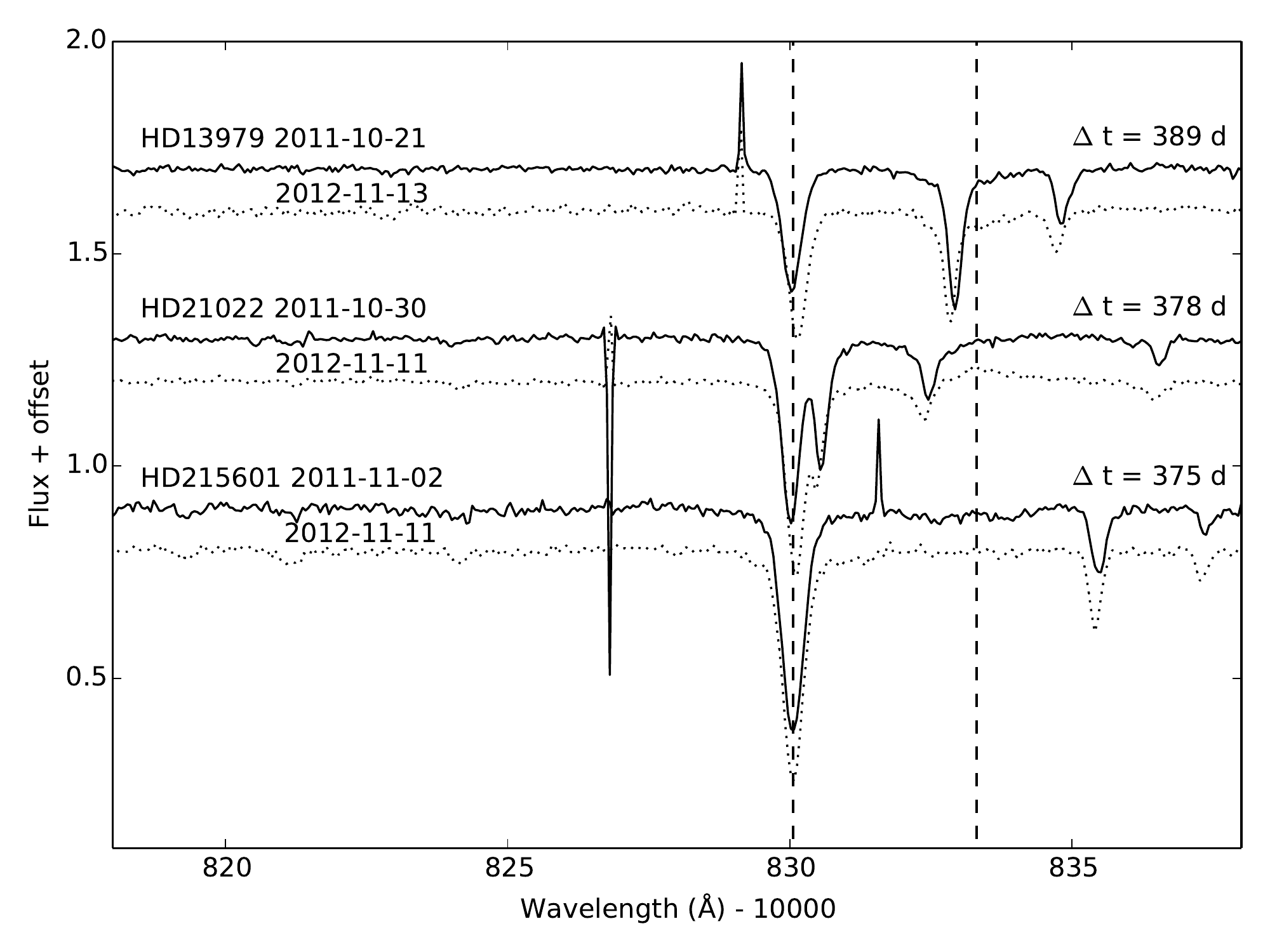}
\caption{CRIRES spectra for HD~13979, HD~21022, and HD~215601 obtained at two different epochs. The 
dates of observations and the timespan between them are indicated for each star. The dashed lines 
mark the positions of the \sii\ (left) and the \hei\ (right) lines. HD~21022 shows the \hei\ line 
in emission in its 2012 spectrum.}
\label{2nd_run}
\end{figure*}

Given that the \hei\ 10830 {\AA} line is associated with activity in cool stars, we also explore its 
variability among a few of the stars in our samples using our own data obtained for another 
program, as well as some measurements available in the literature. CRIRES spectra were obtained 
for HD~13979, HD~21022, and HD~215601 about a year after the observations reported in the present 
work. Using the same reduction process, the spectrum around 10830~{\AA} was extracted, and in 
Figure~\ref{2nd_run} we show the normalized spectra at the two epochs for the three stars. Both 
HD~13979 and HD~215601 show an indistinguishable helium line shape between the two observations. In  
contrast, the \hei\ line in HD~21022 changes from being not detected on the first run (EW smaller 
than 10 m{\AA}) to appearing in emission a year later, with an EW of 50 m{\AA}. 

Finally, we found differences in the \hei\ EW measurements in the literature for two stars in 
our $\omega$~Cen group (see Table~\ref{ew_lit}). For HD~3567, \cite{S12} determine an EW that 
is about half of that derived here, while for HD~140283, \cite{T11} report an EW five times 
larger than ours. For HD~83212, from our field sample, while our data show \hei\ non-detections, 
while observations by \cite{D09} show the line in emission, with an EW of 43.6 m{\AA}.

In summary, out of twelve stars in our full sample with multiple measurements of the \hei\ 10830 
{\AA} line, we find that four present significant variability between separate epochs. 
This suggests that the line should at least be used with caution as an abundance indicator.

\subsection{Barium}\label{barium_disc}

Barium is one of the neutron-capture elements. It is the product of the s-process that takes place 
during the thermal pulses of AGB stars. Therefore, the barium abundance is a proxy for measuring the 
s-process contribution in a stellar system. $\omega$~Centauri has a unique [Ba/Fe]--[Fe/H] distribution 
that does not follow the expected distribution of MW or dwarf spheroidal (dSph) galaxy stars 
\citep[e.g.,][]{G07, M11}. For the most metal-rich population (between $-1.5 \leq$ [Fe/H] $\leq -0.5$ 
dex), $\omega$~Centauri's stars have an extreme overabundance of barium, [Ba/Fe] $\sim$ 0.7 dex, while 
the metal-poor stars experience a sharp increase in [Ba/Fe] with [Fe/H] between $-2.0$ and  
$-1.5$ dex. This is the result of the high contribution of AGB stars to the chemical enrichment 
process that occurs between the formation of the metal-poor and metal-intermediate components 
of the cluster \citep{M11}. To date, only \cite{M12} and \cite{Aquarius14} seem to have identified stars 
that appear to be consistent with an origin in $\omega$~Centauri based on this unique Ba 
abundance pattern.

Figure~\ref{Ba} shows the [Ba/Fe]--[Fe/H] distribution for our target stars, in comparison with MW 
and $\omega$~Centauri stars from previous studies. MW stars predominantly have Ba abundances near 
the solar value, typical of the Galactic halo population. As a whole, neither Kapteyn nor 
$\omega$~Cen group stars follow a trend like the one found for $\omega$~Centauri stars, 
which puts into question any possible association between them. 

However, there are two stars with a marked overabundance of Ba within our samples, compared to the 
other target stars with the same metallicity. CD-62 1346, a target from the Kapteyn group, has the 
most extreme Ba abundance of our sample, [Ba/Fe] $\sim$ 1.6 dex (\citetalias{W10} find [Ba/Fe] = 0.95 for this 
star, also the largest value for their sample). [Ba/Fe] determinations for our Kapteyn group sample 
range from $-0.5$ to 0.3 dex with an average value of [Ba/Fe] = $-0.07$, if CD-62 1346 is excluded. 
The [Ba/Fe] abundances derived by \citetalias{W10} have an average value of 0.21 dex, and 0.16 dex when 
excluding CD-62 1346, both higher than the determinations in the present work. However, the trend 
with metallicity even in \citetalias{W10} is still inconsistent with that of $\omega$~Centauri. 

The other star with an overabundance of Ba is HD~222925, but it belongs to our field control sample 
and we discuss its case further in Section~\ref{peculiar}.

\subsection{$\alpha$-Elements and the Present-day Dwarf Galaxies}

$\omega$~Centauri is thought to be the remnant nucleus of a completely disrupted dwarf galaxy 
\citep{D99, B03, BN06}. Based on that, it seems necessary to address the possibility that the Kapteyn 
and $\omega$~Cen groups originated in the main body of the defunct progenitor galaxy instead of 
the GC itself.

\cite{Carretta10} compared detailed chemical abundances for $\omega$~Centauri RGB stars and 
the stars in the Sagittarius dSph galaxy and its nuclear GC, M54. 
Based on $\feh$ metallicities, the authors concluded that $\omega$~Centauri's stars resemble the 
metallicity distribution found for M54 and Sagittarius dSph together. In particular, the metal-rich 
population of $\omega$~Centauri is consistent with the extent of the metallicity distribution of 
Sagittarius itself. Moreover, the $\alpha$-element abundance for the surrounding nucleus of Sagittarius is 
similar to what is found for the metal-rich stars in $\omega$~Centauri while M54 shows [$\alpha$/Fe] 
abundances in agreement with the metal-poor and metal-intermediate populations of $\omega$~Centauri.

The abundances of three alpha-elements were obtained from our optical spectra (O, Mg, and Ca), 
so we can compare the $\alpha$-element abundances of our target stars with those measured for stars 
in dSph galaxies. Since most of the $\alpha$-abundances in the literature are computed from those 
of Mg and Ca, we use these elements in our own determinations, leaving O aside.

\begin{figure*}[htb!]
\centering
\includegraphics[width=\linewidth]{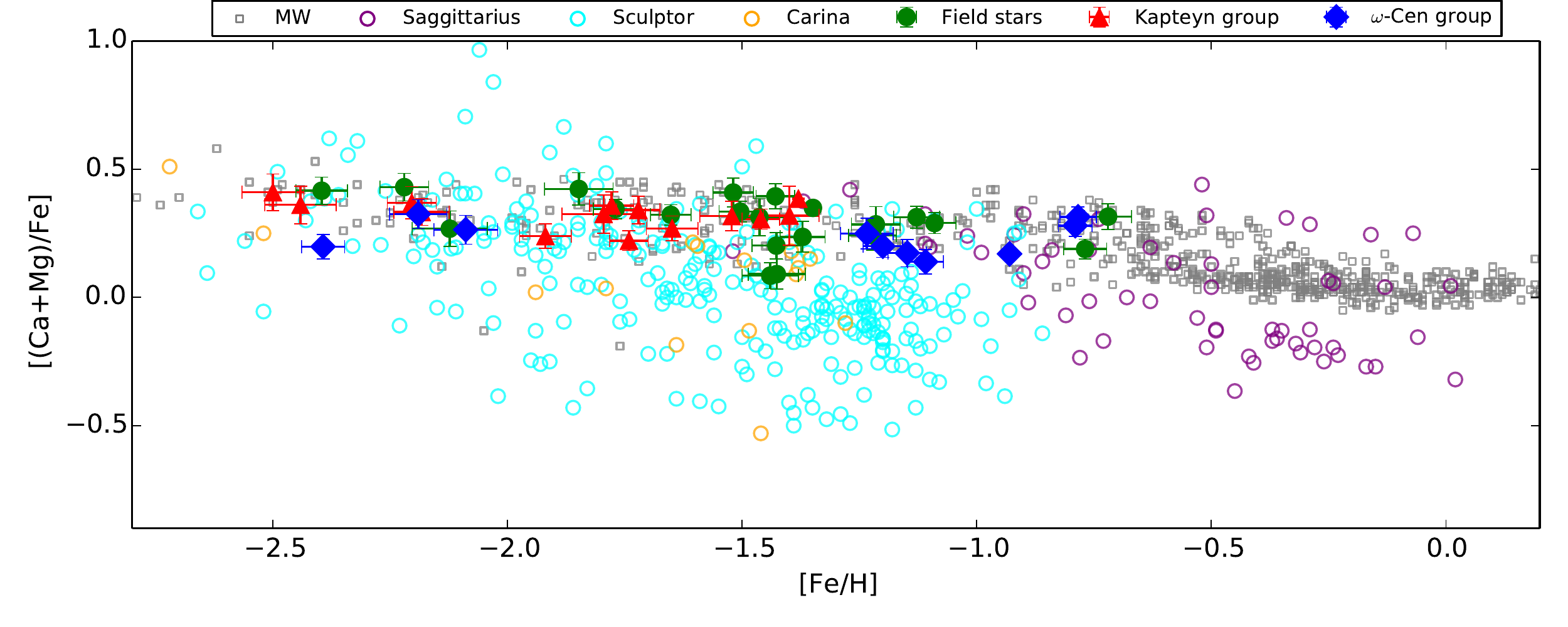}
\caption{[$\alpha$/Fe] abundances for our target stars (same symbols as in Figure~\ref{ONa}) based 
on the Mg and Ca chemical abundances. Gray open squares are field Milky Way stars from both the disk 
and halo \citep{V04}. Open purple circles are for stars in the main body of the Sagittarius dSph galaxy 
\citep{Monaco05, S07, Carretta10}, open cyan circles for stars in the Sculptor dSph galaxy \citep{S03, G05, K09}, 
and open orange circles for stars in the Carina dSph \citep{S03, K08}.}
\label{alpha}
\end{figure*}

Figure~\ref{alpha} shows the [$\alpha$/Fe] abundance versus metallicity for our target stars, 
MW stars, and stars from the Sagittarius, Sculptor, and Carina dSph galaxies. The metallicity 
distributions of both the  Kapteyn and $\omega$~Cen groups do not overlap with that of Sagittarius.
As the GC $\omega$~Centauri appears consistent with the metal-rich stars 
($\feh \geq -1.46$ dex) in Sagittarius, it appears straightforward to discard the idea that the Kapteyn and 
$\omega$~Cen groups' stars have an origin in the defunct host galaxy of $\omega$~Centauri, if such 
an extinct host was comparable to what we see today in Sagittarius.

For Carina and Sculptor, and although they cover the same metallicity range as our target stars, 
the range of [$\alpha$/Fe] variation for stars in those dSphs is very broad and incompatible with our 
stars. In fact, Kapteyn and $\omega$ Cen groups' stars follow, at least qualitatively, the same trend as 
MW field halo stars. At $\feh$ $\leq$ $-1.5$ dex, halo stars \citep{V04} have an average 
[$\alpha$/Fe] abundance of 0.30 with a dispersion of 0.12 dex. The comparison field stars in our 
sample have a very similar average [$\alpha$/Fe] abundance, of 0.30 dex with $\sigma$ = 0.10 dex. 
For Kapteyn group stars, the same average is 0.33 dex ($\sigma$ = 0.06 dex), completely compatible 
with MW halo stars. Stars from the $\omega$~Cen group, on the other hand, have an average 
[$\alpha$/Fe] = 0.23 ($\sigma$ = 0.06 dex), still consistent with the MW halo given the 
dispersion in both distributions, but less so than observed for the Kapteyn group.

Based on the run of $\alpha$-element abundances, the stars of the Kapteyn group do not resemble 
the present-day dSph galaxies close to the MW, including the case of the Sagittarius dwarf, 
thus rejecting $\omega$~Centauri's parent galaxy as the origin for this moving group. The same 
can be said for the stars in the $\omega$~Cen group, but the somewhat lower overall [$\alpha$/Fe] 
among these stars might be indicative of something different and does not allow us to discard 
completely an origin in a dSph galaxy.

\subsection{Chemically peculiar stars}\label{peculiar}

Even though we do not find our samples of the Kapteyn and $\omega$~Cen groups to contain any 
second-generation stars, there are a few stars that appear special on their own, and we examine 
them separately here.

HD~151559, from our field sample, appears as a Na-rich and relatively O-poor star, compared with 
the other stars in our sample. This is one of the most metal-rich stars in the sample, with 
$\feh \sim -0.7$, consistent with the metal-rich population found in $\omega$~Centauri, which is 
mainly composed of second-generation stars. In fact, HD~151559 falls in the same location of Figure~\ref{ONa} 
and of the right panel of Figure~\ref{3panels} where \cite{R12} identified the first field dwarfs with 
O and Na abundances consistent with second-generation stars from GCs. One star out of 45 with both 
O and Na measurements amounts to just in between the 2-3\% of second-generation stars in the 
Galactic field recently found by \cite{Martell11} and \cite{R12}. The \hei\ EW measured for this 
star is below the maximum usually found in the literature for field stars. However, it is slightly 
larger than what is found for dwarfs and subdwarfs with $\feh\ \gtrsim -1.0$ dex \citep{T11}. 
The magnesium and aluminum abundances for HD~151559 (Figure~\ref{MgAl}) are typical of MW halo 
stars. However, as discussed earlier, this alone does not say anything about its origin as a first- 
or second-generation star, since the Mg--Al anticorrelation is evident only in a few massive GCs, like 
$\omega$~Centauri, while other clusters with known different stellar generations are practically 
indistinguishable from the MW halo in its Mg and Al abundance content \citep[see Figure~6 of ][]{C09b}. 
It is the location of HD~151559 in the Na--O diagram that makes this star a firm candidate for a 
second-generation GC star in the Galactic halo.

CD-62 1346 and HD~222925 have an overabundance of Ba and resemble the Ba-rich stars typical of 
$\omega$~Centauri. However, even though their large \hei\ EWs would be expected for helium-enhanced 
stars, their Na and O abundances are similar to field stars, which is not consistent with being 
second-generation stars from $\omega$~Centauri \citep{G11, M11}. Therefore, the large \hei\ EW in these 
cases might be due to chromospheric activity on those stars.

On the other hand, the high [Ba/Fe] values for these stars could be the result of mass transfer from an 
unseen AGB companion which changes the surface abundance of these stars. The most straightforward 
test for this possibility is to search for RV variations in time. For CD-62 1346 a RV of 125.74 
$\pm$0.35 km s${}^{-1}$ is found based on our optical spectrum, which is in excellent agreement with 
the value listed by \cite{B00} of 127.0 $\pm$ 10 km s${}^{-1}$. Despite CD-62 1346 not showing RV 
variations, its $\alpha$ and s-process element abundances are consistent with a CH-rich Ba-strong 
star in a binary system \citep[see][and references therein]{P12}. Therefore, the barium overabundance 
of this star is not representative of the Kapteyn group. HD~222925 is listed with 
$V_{\rm hel}$ = $-$45 $\pm$ 10 km s${}^{-1}$ by \citeauthor{B00} while we found $-$38.64 $\pm$ 0.36 
km s${}^{-1}$. Due to the large uncertainties of the \cite{B00} measurement, it is hard to conclude 
anything for this star. The abundances of yttrium and technetium are desirable to reject mass transfer 
from an AGB companion \citep[][and references therein]{B01, U11, C14}, but this kind of analysis 
is beyond the scope of this study. 

Finally, HD~184266 has the largest value ($>$ 400 m{\AA}) for the \hei\ EW of the whole sample. 
This star from our field sample is indistinguishable from the field halo stars with the same 
metallicity in its O, Na and Ba abundances. From visual inspection of its optical 
spectrum obtained with MIKE, the core of the chromospheric Ca II lines is not found in emission, 
undermining the possibility of chromospheric activity as being responsible for the high \hei\ EW 
\citep[see][]{P11}. The same is found for HD~222925, the star with the second largest \hei\ 
EW ($\sim$~230 m{\AA}), also from our field star sample. A more detailed chromospheric/activity 
study of these stars is needed in order to determine whether these EWs are consistent with 
enhanced helium abundances\footnote{Although HD~222925 appears as an Ap star in the literature 
\citep[][and the subsequent compilations of HD stars]{H75}, the study of \cite{R09} classified 
it as an F8 star.}.

\section{Kinematics}\label{sec:kine}

\begin{deluxetable*}{cccccccccc}
\tablecaption{Distances and Galactic Velocities for our Target Stars.\label{kine}}
\tablecolumns{8}
\tablewidth{0pt}
\tablehead{\colhead{Star} & \colhead{Distance} & \colhead{$U$} & \colhead{$\sigma_U$} & \colhead{$V$} & \colhead{$\sigma_V$} & \colhead{$W$} & \colhead{$\sigma_W$} \\
 \colhead{} & \colhead{(pc)} & \colhead{(km s${}^{-1}$)} & \colhead{(km s${}^{-1}$)} & \colhead{(km s${}^{-1}$)}  & \colhead{(km s${}^{-1}$)} & \colhead{(km s${}^{-1}$)} & \colhead{(km s${}^{-1}$)} } 
\startdata
\multicolumn{8}{c}{\textbf{Kapteyn Group Stars}}\\
\hline
CD-30 1121 &  842 &   12  &  6 & -191 & 33 &  -28  & 12 \\ 
CD-62 1346 &  243 &  -78  &  4 & -149 & 22 &  -34  &  8 \\ 
HD 110621  &  191 &   47  & 34 & -285 & 22 &   55  &  4 \\ 
HD 111721  &  188 &   69  & 17 & -322 & 63 & -163  & 37 \\ 
HD 13979   &  444 &  -21  &  5 & -120 & 23 &  -27  &  3 \\ 
HD 181007  &  410 &  -92  & 17 & -317 & 64 &  -55  & 13 \\ 
HD 181743  &  120 &   53  & 18 & -449 & 90 &  -80  & 15 \\ 
HD 186478  & 1003 &  -173 & 27 & -363 & 76 &  -65  & 14 \\ 
HD 188031  &  181 &  174  & 12 & -351 & 73 &    7  & 14 \\ 
HD 193242  &  326 &   42  & 10 & -185 & 29 &   46  &  5 \\ 
HD 208069  &  257 &  162  & 15 & -282 & 51 &   20  & 23 \\ 
HD 21022   & 1285 &  -31  & 14 & -300 & 51 &    9  & 21 \\ 
HD 215601  &  263 &   11  &  1 & -194 & 39 &   10  &  6 \\ 
HD 215801  &  150 &   16  &  4 & -198 & 43 &  123  &  8 \\ 
\hline
\multicolumn{8}{c}{\textbf{$\mathbf{\omega}$~Cen group Stars}}\\
\hline
BD+02 3375 &  147 &  362 &  7 & -263  & 19 &  129  & 48  \\ 
CD-61 0282 &  122 & -325 & 53 & -316  & 41 &   31  & 41  \\ 
HD 113083  &   40 &  -55 & 13 & -215  & 14 &  112  &  6  \\ 
HD 121004  &   53 &  -85 & 18 & -241  & 15 &  102  &  6  \\ 
HD 140283  &   26 &  180 & 10 & -104  & 24 &  -25  & 12  \\ 
HD 148816  &   29 &  -60 & 18 & -185  & 35 &  -56  &  7  \\ 
HD 193901  &   54 &  150 &  4 & -285  & 47 & -103  & 39  \\ 
HD 194598  &   61 &   59 & 15 & -280  & 19 &  -33  & 22  \\ 
HD 3567    &   99 & -138 & 24 & -219  & 42 &  -31  & 17  \\ 
HD 84937   &   86 & -251 & 47 & -250  & 52 &    0  &  1  \\ 
\hline
\multicolumn{8}{c}{\textbf{Field Stars}}\\
\hline
HD 102200 &   94 & -106 &  8   & -136 &  1   &   11  &  8 \\ 
HD 116064 &   87 &  163 & 51   & -261 & 33   &  138  & 15 \\ 
HD 128279 &  133 &   -7 & 11   &  -65 & 21   & -214  & 37 \\ 
HD 134439 &   32 & -318 & 13   & -547 & 97   &  -84  & 54 \\ 
HD 134440 &   34 & -322 & 13   & -577 & 103  & -101  & 57 \\ 
HD 142948 &  131 &  -22 &  3   &  -32 &  4   &   -1  &  2 \\ 
HD 145417 &   15 &   46 & 13   &  -94 & 19   &  -24  &  6 \\ 
HD 151559 &  265 &  -23 &  1   &   21 &  4   &   39  &  6 \\ 
HD 17072  &   38 &  -19 &  1   &  -45 &  2   &  -31  &  1 \\ 
HD 184266 &   54 &  299 &  1   & -160 & 7    &   58  & 10 \\ 
HD 190287 &  142 & -145 &  2   &  -95 & 23   &  -67  &  1 \\ 
HD 199289 &   49 &   27 &  7   &  -58 & 13   &  -15  &  5 \\ 
HD 211998 &   30 &  138 & 33   & -129 & 24   &  -61  &  9 \\ 
HD 219617 &   58 & -278 & 53   & -225 & 47   &  -37  &  6 \\ 
HD 221580 &  160 &   60 & 13   &  -71 & 16   &   18  &  1 \\ 
HD 222434 & 1120 & -108 & 19   & -214 & 44   &   17  &  5 \\ 
HD 222925 &  115 &   69 & 12   &  -57 & 16   &   46  &  2 \\ 
HD 23798  &  995 &   57 & 7    & -105 & 14   &   -5  & 12 \\ 
HD 26169  &  351 &  196 & 39   & -111 & 29   &  119  & 18 \\ 
HD 83212  &  563 &   19 &  2   & -119 & 6    &  -15  & 13 \\ 
HD 9051   &  485 &   51 & 15   &  -96 & 21   &   94  &  3 \\ 
 \enddata
\end{deluxetable*}

The Kapteyn moving group was originally selected on the basis of kinematics (proper motions and 
Galactic velocities), consistent with retrograde Galactic orbits \citep{E77}. Meanwhile, the 
$\omega$~Cen group was kinematically selected by \cite{M05} considering its broad, symmetric, 
double-peaked distribution of Galactocentric RVs and low vertical velocities 
consistent with the expectations for tidal debris from accreted satellite galaxies.

Despite the fact that all our targets have \textit{Hipparcos} trigonometric parallax measurements, 
the adopted distances were determined based on the photometric relations derived by \cite{B00}. 
\textit{Hipparcos} distance estimates were discarded because most of our targets are distant stars, 
hence its parallax measurements are very inaccurate, with relative errors 
$\sigma_{\pi_{\rm HIP}}/\pi_{\rm HIP}$ larger than 15\%. 

\begin{figure*}[htb!]
\centering
\includegraphics[width=0.75\linewidth]{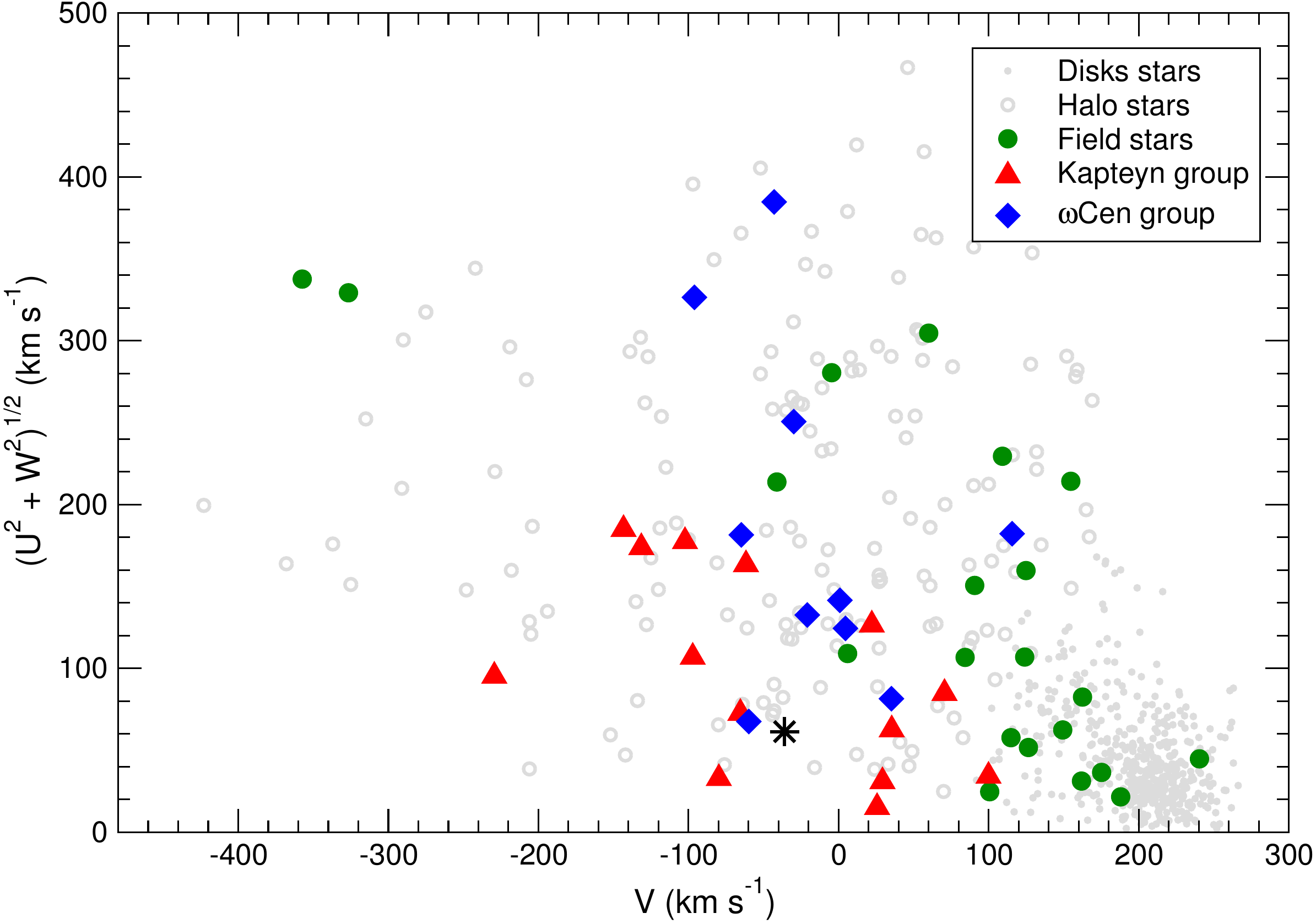}
\caption{Toomre diagram showing the stars from our target groups (same symbols as in 
Figure~\ref{ONa}). The black star represents the position of $\omega$~Centauri, based on its 
present velocity components. Thick and thin disk stars (gray dots) and halo stars (open gray 
circles) from \cite{V04} are shown for comparison. \label{fig:toomre}}
\end{figure*}

\cite{B00} provide relations for the $V$-band absolute magnitudes, based on the stellar type 
(e.g., giants, dwarfs, AGB stars), which are linear functions of the metallicity and dereddened 
$(B-V)_0$ color of each star. The stellar types of our stars were determined spectroscopically 
from our optical spectra, and the metallicity values adopted were those derived 
in the present work (see Table~\ref{targets}). The $(B-V)$ colors and the 
color excess $E(B-V)$ were adopted from the catalogs of \cite{B00} and \cite{G03}. 
The derived distances are presented in Column 2 of Table~\ref{kine}.

The galactic space velocities were calculated using these distances, our derived RVs 
from the optical spectra, and the tabulated proper motions measured by \cite{vL07}. 
Table~\ref{kine} shows the galactic velocities $(U, V, W)$, where $U$ is defined as positive toward 
the Galactic anticenter, $V$ is in the direction of the Sun's rotation, and $W$ points toward the 
north Galactic pole. The velocities were corrected for the solar motion in the LSR, adopting 
$(U, V, W)_\odot$ = $(-10.00, 5.25, 7.17)$ from \cite{D98}. The position and motion of the Sun 
were also corrected, adopting $X_\odot$ = 8.0 kpc and $V_\odot$ =~220 km s$^{-1}$. The errors listed 
for the Galactic velocities in Table~\ref{kine} were determined by error propagation, adopting an 
error of 20\% for the distances \citep[mean errors of the photometric distances derived by][]{B00}, 
using the formal errors for the proper motions as listed by \cite{vL07}, and our derived errors for 
the RVs.

Figure \ref{fig:toomre} shows the Toomre diagram for stars from the Kapteyn and $\omega$~Cen groups. 
For comparison, we also show halo stars (open gray circles) and disk stars (gray dots) from \cite{V04}. 
All stars from both groups are more probably associated with the halo population.

\begin{figure*}[htb!]
\centering
\includegraphics[width=0.75\linewidth]{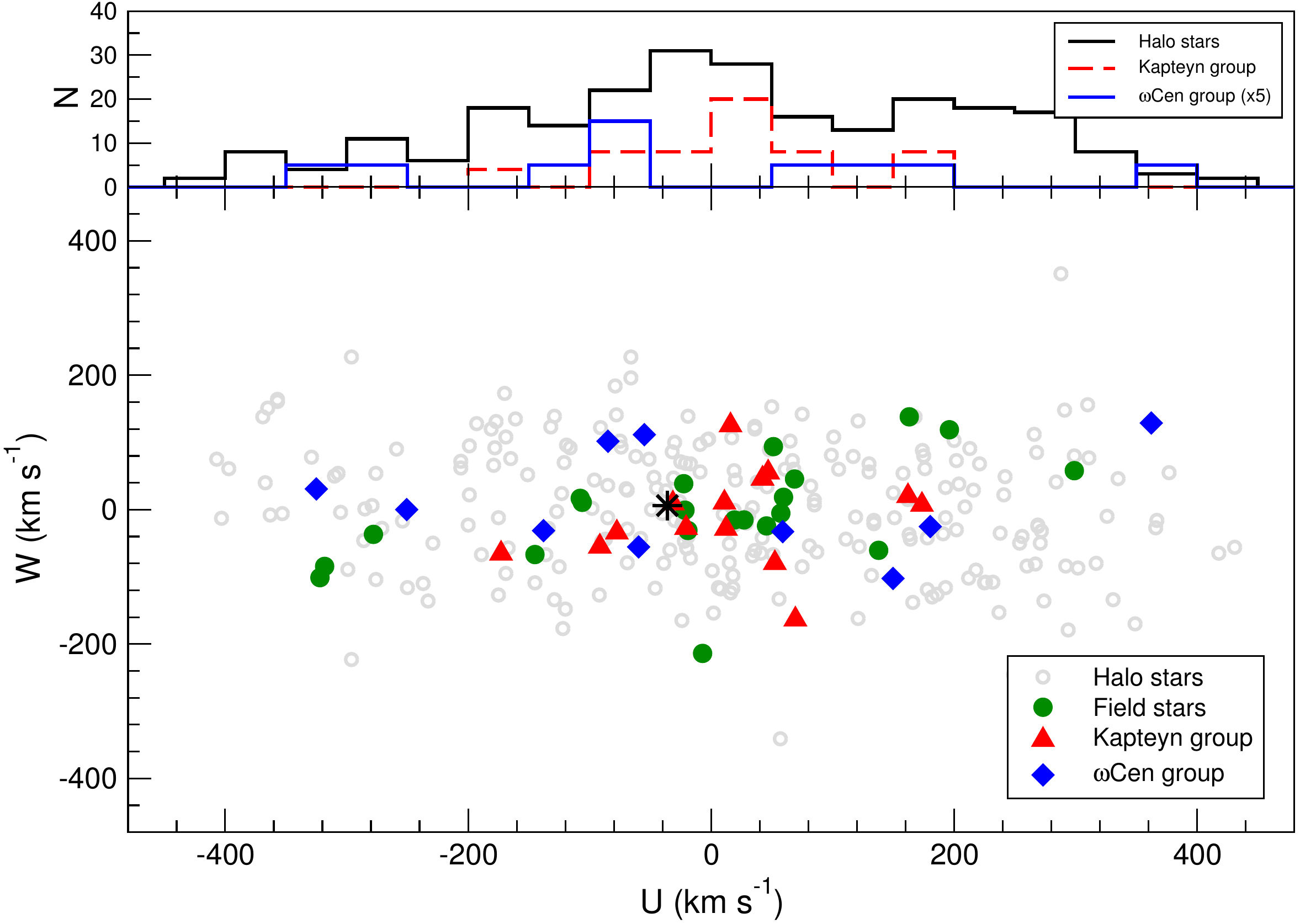}
\caption{Vertical $W$ versus radial $U$ Galactic velocities for the stars from the Kapteyn group, 
$\omega$~Cen-group, and field stars (same symbols as in Figure~\ref{ONa}). The black star 
represents the present velocity components of $\omega$~Centauri. Metal-poor stars with angular 
momentum $-1000 < J_z < 0$  kpc km s$^{-1}$ from the \cite{B00} catalogue are shown as grey open 
circles. \label{fig:UW}}
\end{figure*}

Following \cite{M05}, we choose to analyze the Galactocentric radial $U$-velocity distribution of 
Kapteyn and $\omega$~Cen group stars because relatively symmetric, double-peaked Galactic RV 
distributions are an expected sign of a past accretion event. This simple signature does 
not rely on the computation of energies or other dynamical variables, which can introduce unnecessary 
complications or problems of interpretation. \citetalias{W10} consider the angular momentum--energy plane to 
select the members of the Kapteyn group that are kinematically coherent, within $\pm$1$\sigma$, with 
the theoretical prediction of \cite{D02} for $\omega$~Centauri candidate stars in the halo. We prefer 
not to use this approach since it is strongly dependent on the distance errors, the adopted LSR, 
and a largely subjective decision on how far from the present $\omega$ Centauri orbital elements 
to allow candidate stars. Figure~\ref{fig:UW} shows the vertical $W$ versus radial $U$ velocity for 
stars in the Kapteyn (red circles) and $\omega$~Cen (blue circles) groups. This figure also shows 
metal-poor halo stars with angular momentum $-1000 < J_z < 0$  kpc km s$^{-1}$ from the catalog of \cite{B00}.
The U distribution of the 14 stars from the Kapteyn group (red dashed-line histogram, 
multiplied by four for clarity) does not differ significantly from that of the stars in the 
\citeauthor{B00} compilation (gray solid-line histogram). This argues against a common origin of 
the Kapteyn group and the accretion of the $\omega$~Centauri parent object. In the case of the 
$\omega$~Cen group, the U distribution of the ten stars observed (blue solid-line histogram, 
multiplied by five) is somewhat different. Instead of having a peak at $U$ = 0 km s$^{-1}$, the 
distribution presents peaks at positive and negative $U$ values, consistent with the expected 
U-velocity distribution for accreted stars in the Galaxy \citep{M05}. Clearly, this signal 
is low, but still suggestive of an extragalactic origin for the $\omega$~Cen group, 
as previously claimed by \cite{M05}.

\section{Conclusions}\label{sec:conclusions}

Based on newly determined chemical abundances of 14 stars belonging to the Kapteyn moving group, 
we find no evidence that this group could be identified as tidal debris from $\omega$~Centauri, 
as claimed elsewhere. Neither does the kinematically selected $\omega$~Cen group show any chemical 
signatures resembling the abundance patterns seen among $\omega$ Centauri's stars.

First, stars in both groups have O and Na abundances indistinguishable from the field halo stars, 
and there is no evidence for an anticorrelation between O and Na, as expected for GC stars. The same 
applies to the Mg and Al abundances. If the Kapteyn and $\omega$~Cen moving groups are stripped 
stars from $\omega$~Centauri, four and three second-generation stars, respectively, are expected to be 
in our sample. Based on the absence of second-generation chemical abundances for the stars in the Kapteyn 
and $\omega$~Cen groups, we discard a possible connection between the two moving groups and $\omega$~Centauri.

Second, stars in the Kapteyn group have a mean value of [Ba/Fe] = $-0.07$ dex (excluding 
CD-62~1346, the most Ba-rich star in the sample), consistent with the mean barium abundance of 
disk and halo stars. Moreover, the specific trend between Ba and metallicity seen in 
$\omega$~Centauri's stars is not followed by either the Kapteyn group or the $\omega$~Cen group. 
The barium abundances are consistent with field stars, contrary to the $\omega$~Centauri tidal debris 
candidate stars found by \cite{M12}.

Third, stars in the Kapteyn and $\omega$ Cen groups have EW values for the chromospheric \hei\ 
10830 {\AA} line that are indistinguishable from non-active field stars. Thus there is no hint of 
helium enhancement among the Kapteyn group stars, which would have been a very suggestive link 
with the helium-enhanced second-generation stars that $\omega$~Centauri hosts \citep[][and references 
therein]{D11, D13}. We were able to track non-negligible levels of variability of the \hei\ 10830 
{\AA} line for a small number of our targets, which suggests that its use as helium abundance indicator 
should be treated with caution.

Fourth, the kinematic analysis of the $U$-velocity distribution of the Kapteyn group does not 
support the accretion scenario. As \cite{M05} pointed out, tidal debris from 
satellite galaxies are expected to show a bimodal distribution of $U$ velocities, which is not 
seen in the Kapteyn group. A bimodal, symmetric distribution of $U$ velocities can be 
identified among the stars in the $\omega$~Cen group, still allowing these stars as to be considered 
as candidates to be accreted onto the MW in the past.

It thus may appear that the only possibility remaining for the Kapteyn group to still be associated 
to $\omega$~Centauri's parent system is that all the stars in our sample were first-generation stars 
selectively stripped from the cluster (i.e., the primordial component, which is indistinguishable 
from the halo field stars at a given metallicity). However this possibility is unlikely (see 
Section~\ref{ona_disc}). Even assuming that this is the case, the Ba--$\feh$ pattern rejects this 
possible connection, since even first-generation stars in $\omega$~Centauri are different from the 
MW halo field stars in their barium content (see Section~\ref{barium_disc}).

Based on all of the above, we conclude that neither the Kapteyn group nor the $\omega$~Cen group consists  
of stars stripped from $\omega$~Centauri, as suggested elsewhere \citep{M05, W10}. Our chemical study 
is not consistent with stars in either of these groups, when considered as a group, 
having an origin in any GC, as their chemical abundances are indistinguishable from 
the MW halo stars.

Finally, the $\alpha$-element abundances of the Kapteyn group strongly resemble the abundances 
found for MW halo stars. dSph galaxies have lower values of [$\alpha$/Fe] compared to our Galaxy, 
which is not seen in any of the stars in the Kapteyn group. Our measurements therefore do not support 
the idea that this moving group could originate in the defunct galaxy progenitor of $\omega$~Centauri, 
but rather reinforce the hypothesis that those stars are chemically indistinguishable from the field 
MW halo stars.

The $\omega$~Cen group, in contrast, have slightly lower [$\alpha$/Fe] abundances, when 
compared to the MW stars in the same range of metallicity, just like stars in dSph galaxies. Therefore, 
some of the stars in this group could have originated in a progenitor that was accreted in 
the MW. Moreover, the $\omega$~Cen group stars could remain as a coherent group because their $U$-velocity 
distribution is different from what is observed in the MW halo.

\begin{acknowledgements} 
\textit{Acknowledgements: We thank Elena Valenti for valuable help on the wavelength calibration of 
the CRIRES spectra. We also thank Ian Thompson for obtaining MIKE spectra for stars that were not 
observable during our own Magellan runs. Support for this project is provided by the FONDECYT 
grant \#1130373 (C.N., J.C.); by  BASAL PFB-06 ``Centro de Astronom\'ia y Tecnolog\'ias Afines''; 
and by the Ministry for the Economy, Development, and Tourism's Programa Iniciativa 
Cient\'ifica Milenio through grant IC 120009, awarded to the Millennium Institute of Astrophysics 
(MAS). C.N. acknowledges additional support from CONICYT-PCHA/Doctorado Nacional grant 2015-21151643.}
\end{acknowledgements}

\bibliography{biblio}

\begin{thebibliography}{104}
\expandafter\ifx\csname natexlab\endcsname\relax\def\natexlab#1{#1}\fi

\bibitem[{{Altmann} {et~al.}(2005){Altmann}, {Catelan}, \&
  {Zoccali}}]{Altmann05}
{Altmann}, M., {Catelan}, M., \& {Zoccali}, M. 2005, \aap, 439, L5

\bibitem[{{Asplund} \& {Garc{\'{\i}}a P{\'e}rez}(2001)}]{A01}
{Asplund}, M. \& {Garc{\'{\i}}a P{\'e}rez}, A.~E. 2001, A\&A, 372, 601

\bibitem[{{Bedin} {et~al.}(2004){Bedin}, {Piotto}, {Anderson}, {Cassisi},
  {King}, {Momany}, \& {Carraro}}]{B04}
{Bedin}, L.~R., {Piotto}, G., {Anderson}, J., {et~al.} 2004, ApJ, 605, L125

\bibitem[{{Beers} {et~al.}(2000){Beers}, {Chiba}, {Yoshii}, {Platais},
  {Hanson}, {Fuchs}, \& {Rossi}}]{B00}
{Beers}, T.~C., {Chiba}, M., {Yoshii}, Y., {et~al.} 2000, AJ, 119, 2866

\bibitem[{{Bekki} \& {Freeman}(2003)}]{B03}
{Bekki}, K. \& {Freeman}, K.~C. 2003, MNRAS, 346, L11

\bibitem[{{Bekki} \& {Norris}(2006)}]{BN06}
{Bekki}, K. \& {Norris}, J.~E. 2006, \apjl, 637, L109

\bibitem[{{Bellini} {et~al.}(2010){Bellini}, {Bedin}, {Piotto}, {Milone},
  {Marino}, \& {Villanova}}]{B10}
{Bellini}, A., {Bedin}, L.~R., {Piotto}, G., {et~al.} 2010, AJ, 140, 631

\bibitem[{{Bellini} {et~al.}(2009){Bellini}, {Piotto}, {Bedin}, {King},
  {Anderson}, {Milone}, \& {Momany}}]{B09}
{Bellini}, A., {Piotto}, G., {Bedin}, L.~R., {et~al.} 2009, A\&A, 507, 1393

\bibitem[{{Belokurov} {et~al.}(2006){Belokurov}, {Evans}, {Irwin}, {Hewett}, \&
  {Wilkinson}}]{B06a}
{Belokurov}, V., {Evans}, N.~W., {Irwin}, M.~J., {Hewett}, P.~C., \&
  {Wilkinson}, M.~I. 2006, ApJ, 637, L29

\bibitem[{{Bensby} {et~al.}(2013){Bensby}, {Yee}, {Feltzing}, {Johnson},
  {Gould}, {Cohen}, {Asplund}, {Mel{\'e}ndez}, {Lucatello}, {Han}, {Thompson},
  {Gal-Yam}, {Udalski}, {Bennett}, {Bond}, {Kohei}, {Sumi}, {Suzuki}, {Suzuki},
  {Takino}, {Tristram}, {Yamai}, \& {Yonehara}}]{B13}
{Bensby}, T., {Yee}, J.~C., {Feltzing}, S., {et~al.} 2013, A\&A, 549, A147

\bibitem[{{Breckinridge} \& {Hall}(1973)}]{B73}
{Breckinridge}, J.~B. \& {Hall}, D.~N.~B. 1973, \solphys, 28, 15

\bibitem[{{Busso} {et~al.}(2001){Busso}, {Gallino}, {Lambert}, {Travaglio}, \&
  {Smith}}]{B01}
{Busso}, M., {Gallino}, R., {Lambert}, D.~L., {Travaglio}, C., \& {Smith},
  V.~V. 2001, ApJ, 557, 802

\bibitem[{{Carretta} {et~al.}(2009{\natexlab{a}}){Carretta}, {Bragaglia},
  {Gratton}, \& {Lucatello}}]{C09}
{Carretta}, E., {Bragaglia}, A., {Gratton}, R., \& {Lucatello}, S.
  2009{\natexlab{a}}, A\&A, 505, 139

\bibitem[{{Carretta} {et~al.}(2007){Carretta}, {Bragaglia}, {Gratton},
  {Catanzaro}, {Leone}, {Sabbi}, {Cassisi}, {Claudi}, {D'Antona}, {Fran{\c
  c}ois}, {James}, \& {Piotto}}]{C07}
{Carretta}, E., {Bragaglia}, A., {Gratton}, R.~G., {et~al.} 2007, \aap, 464,
  939

\bibitem[{{Carretta} {et~al.}(2010{\natexlab{a}}){Carretta}, {Bragaglia},
  {Gratton}, {Lucatello}, {Bellazzini}, {Catanzaro}, {Leone}, {Momany},
  {Piotto}, \& {D'Orazi}}]{Carretta10}
{Carretta}, E., {Bragaglia}, A., {Gratton}, R.~G., {et~al.} 2010{\natexlab{a}},
  \apjl, 714, L7

\bibitem[{{Carretta} {et~al.}(2009{\natexlab{b}}){Carretta}, {Bragaglia},
  {Gratton}, {Lucatello}, {Catanzaro}, {Leone}, {Bellazzini}, {Claudi},
  {D'Orazi}, {Momany}, {Ortolani}, {Pancino}, {Piotto}, {Recio-Blanco}, \&
  {Sabbi}}]{C09b}
{Carretta}, E., {Bragaglia}, A., {Gratton}, R.~G., {et~al.} 2009{\natexlab{b}},
  \aap, 505, 117

\bibitem[{{Carretta} {et~al.}(2010{\natexlab{b}}){Carretta}, {Bragaglia},
  {Gratton}, {Recio-Blanco}, {Lucatello}, {D'Orazi}, \&
  {Cassisi}}]{Carretta10c}
{Carretta}, E., {Bragaglia}, A., {Gratton}, R.~G., {et~al.} 2010{\natexlab{b}},
  \aap, 516, A55

\bibitem[{{Casey} {et~al.}(2014{\natexlab{a}}){Casey}, {Keller}, {Alves-Brito},
  {Frebel}, {Da Costa}, {Karakas}, {Yong}, {Schlaufman}, {Jacobson}, {Yu}, \&
  {Fishlock}}]{Aquarius14}
{Casey}, A.~R., {Keller}, S.~C., {Alves-Brito}, A., {et~al.}
  2014{\natexlab{a}}, MNRAS, 443, 828

\bibitem[{{Casey} {et~al.}(2014{\natexlab{b}}){Casey}, {Keller}, {Da Costa},
  {Frebel}, \& {Maunder}}]{C14}
{Casey}, A.~R., {Keller}, S.~C., {Da Costa}, G., {Frebel}, A., \& {Maunder}, E.
  2014{\natexlab{b}}, ApJ, 784, 19

\bibitem[{{Cayrel}(1988)}]{C88}
{Cayrel}, R. 1988, in IAU Symposium, Vol. 132, The Impact of Very High S/N
  Spectroscopy on Stellar Physics, ed. G.~{Cayrel de Strobel} \& M.~{Spite},
  345

\bibitem[{{Chen} {et~al.}(2000){Chen}, {Nissen}, {Zhao}, {Zhang}, \&
  {Benoni}}]{C00}
{Chen}, Y.~Q., {Nissen}, P.~E., {Zhao}, G., {Zhang}, H.~W., \& {Benoni}, T.
  2000, A\&As, 141, 491

\bibitem[{{D'Antona} {et~al.}(2005){D'Antona}, {Bellazzini}, {Caloi}, {Pecci},
  {Galleti}, \& {Rood}}]{D05}
{D'Antona}, F., {Bellazzini}, M., {Caloi}, V., {et~al.} 2005, \apj, 631, 868

\bibitem[{{Dehnen} \& {Binney}(1998)}]{D98}
{Dehnen}, W. \& {Binney}, J.~J. 1998, MNRAS, 298, 387

\bibitem[{{Dinescu}(2002)}]{D02}
{Dinescu}, D.~I. 2002, in Astronomical Society of the Pacific Conference
  Series, Vol. 265, Omega Centauri, A Unique Window into Astrophysics, ed.
  F.~{van Leeuwen}, J.~D. {Hughes}, \& G.~{Piotto}, 365

\bibitem[{{Dinescu} {et~al.}(1999){Dinescu}, {Girard}, \& {van Altena}}]{D99}
{Dinescu}, D.~I., {Girard}, T.~M., \& {van Altena}, W.~F. 1999, AJ, 117, 1792

\bibitem[{{Dupree} \& {Avrett}(2013)}]{D13}
{Dupree}, A.~K. \& {Avrett}, E.~H. 2013, ApJ, 773, L28

\bibitem[{{Dupree} {et~al.}(2009){Dupree}, {Smith}, \& {Strader}}]{D09}
{Dupree}, A.~K., {Smith}, G.~H., \& {Strader}, J. 2009, AJ, 138, 1485

\bibitem[{{Dupree} {et~al.}(2011){Dupree}, {Strader}, \& {Smith}}]{D11}
{Dupree}, A.~K., {Strader}, J., \& {Smith}, G.~H. 2011, ApJ, 728, 155

\bibitem[{{Edl{\'e}n}(1953)}]{E53}
{Edl{\'e}n}, O.~J. 1953, J. Opt. Soc. Am., 43, 339

\bibitem[{{Eggen}(1977)}]{E77}
{Eggen}, O.~J. 1977, ApJ, 215, 812

\bibitem[{{Eggen}(1978)}]{E78}
{Eggen}, O.~J. 1978, ApJ, 221, 881

\bibitem[{{Eggen}(1996)}]{E96}
{Eggen}, O.~J. 1996, AJ, 112, 1595

\bibitem[{{Fabbian} {et~al.}(2009){Fabbian}, {Asplund}, {Barklem}, {Carlsson},
  \& {Kiselman}}]{F09}
{Fabbian}, D., {Asplund}, M., {Barklem}, P.~S., {Carlsson}, M., \& {Kiselman},
  D. 2009, A\&A, 500, 1221

\bibitem[{{Freeman} \& {Bland-Hawthorn}(2002)}]{F02}
{Freeman}, K. \& {Bland-Hawthorn}, J. 2002, ARA\&A, 40, 487

\bibitem[{{Freeman}(1993)}]{F93}
{Freeman}, K.~C. 1993, in Astronomical Society of the Pacific Conference
  Series, Vol.~48, The Globular Cluster-Galaxy Connection, ed. G.~H. {Smith} \&
  J.~P. {Brodie}, 27

\bibitem[{{Geisler} {et~al.}(2005){Geisler}, {Smith}, {Wallerstein},
  {Gonzalez}, \& {Charbonnel}}]{G05}
{Geisler}, D., {Smith}, V.~V., {Wallerstein}, G., {Gonzalez}, G., \&
  {Charbonnel}, C. 2005, \aj, 129, 1428

\bibitem[{{Geisler} {et~al.}(2007){Geisler}, {Wallerstein}, {Smith}, \&
  {Casetti-Dinescu}}]{G07}
{Geisler}, D., {Wallerstein}, G., {Smith}, V.~V., \& {Casetti-Dinescu}, D.~I.
  2007, PASP, 119, 939

\bibitem[{{Gratton} {et~al.}(2004){Gratton}, {Sneden}, \& {Carretta}}]{G04}
{Gratton}, R., {Sneden}, C., \& {Carretta}, E. 2004, ARA\&A, 42, 385

\bibitem[{{Gratton} {et~al.}(2012){Gratton}, {Carretta}, \& {Bragaglia}}]{G12}
{Gratton}, R.~G., {Carretta}, E., \& {Bragaglia}, A. 2012, A\&ARv, 20, 50

\bibitem[{{Gratton} {et~al.}(2003){Gratton}, {Carretta}, {Claudi}, {Lucatello},
  \& {Barbieri}}]{G03}
{Gratton}, R.~G., {Carretta}, E., {Claudi}, R., {Lucatello}, S., \& {Barbieri},
  M. 2003, A\&A, 404, 187

\bibitem[{{Gratton} {et~al.}(2011){Gratton}, {Johnson}, {Lucatello}, {D'Orazi},
  \& {Pilachowski}}]{G11}
{Gratton}, R.~G., {Johnson}, C.~I., {Lucatello}, S., {D'Orazi}, V., \&
  {Pilachowski}, C. 2011, A\&A, 534, A72

\bibitem[{{Hinkle} {et~al.}(1995){Hinkle}, {Wallace}, \& {Livingston}}]{H95}
{Hinkle}, K., {Wallace}, L., \& {Livingston}, W. 1995, PASP, 107, 1042

\bibitem[{{Houk} \& {Cowley}(1975)}]{H75}
{Houk}, N. \& {Cowley}, A.~P. 1975, {University of Michigan Catalogue of
  two-dimensional spectral types for the HD stars. Volume I. Declinations -90\_
  to -53\_f0.}

\bibitem[{{Ibata} {et~al.}(1994){Ibata}, {Gilmore}, \& {Irwin}}]{I94}
{Ibata}, R.~A., {Gilmore}, G., \& {Irwin}, M.~J. 1994, \nat, 370, 194

\bibitem[{{Ibata} {et~al.}(1995){Ibata}, {Gilmore}, \& {Irwin}}]{I95}
{Ibata}, R.~A., {Gilmore}, G., \& {Irwin}, M.~J. 1995, MNRAS, 277, 781

\bibitem[{{Ishigaki} {et~al.}(2013){Ishigaki}, {Aoki}, \& {Chiba}}]{I13}
{Ishigaki}, M.~N., {Aoki}, W., \& {Chiba}, M. 2013, ApJ, 771, 67

\bibitem[{{Johnson} \& {Pilachowski}(2010)}]{JP10}
{Johnson}, C.~I. \& {Pilachowski}, C.~A. 2010, ApJ, 722, 1373

\bibitem[{{Kaeufl} {et~al.}(2004){Kaeufl}, {Ballester}, {Biereichel},
  {Delabre}, {Donaldson}, {Dorn}, {Fedrigo}, {Finger}, {Fischer}, {Franza},
  {Gojak}, {Huster}, {Jung}, {Lizon}, {Mehrgan}, {Meyer}, {Moorwood}, {Pirard},
  {Paufique}, {Pozna}, {Siebenmorgen}, {Silber}, {Stegmeier}, \&
  {Wegerer}}]{K04}
{Kaeufl}, H.-U., {Ballester}, P., {Biereichel}, P., {et~al.} 2004, in Society
  of Photo-Optical Instrumentation Engineers (SPIE) Conference Series, Vol.
  5492, Ground-based Instrumentation for Astronomy, ed. A.~F.~M. {Moorwood} \&
  M.~{Iye}, 1218--1227

\bibitem[{{Kirby} {et~al.}(2009){Kirby}, {Guhathakurta}, {Bolte}, {Sneden}, \&
  {Geha}}]{K09}
{Kirby}, E.~N., {Guhathakurta}, P., {Bolte}, M., {Sneden}, C., \& {Geha}, M.~C.
  2009, \apj, 705, 328

\bibitem[{{Koch} {et~al.}(2008){Koch}, {Grebel}, {Gilmore}, {Wyse}, {Kleyna},
  {Harbeck}, {Wilkinson}, \& {Wyn Evans}}]{K08}
{Koch}, A., {Grebel}, E.~K., {Gilmore}, G.~F., {et~al.} 2008, \aj, 135, 1580

\bibitem[{{Lee} {et~al.}(2009){Lee}, {Kang}, {Lee}, \& {Lee}}]{L09}
{Lee}, J.-W., {Kang}, Y.-W., {Lee}, J., \& {Lee}, Y.-W. 2009, Nature, 462, 480

\bibitem[{{Majewski} {et~al.}(2012){Majewski}, {Nidever}, {Smith}, {Damke},
  {Kunkel}, {Patterson}, {Bizyaev}, \& {Garc{\'{\i}}a P{\'e}rez}}]{M12}
{Majewski}, S.~R., {Nidever}, D.~L., {Smith}, V.~V., {et~al.} 2012, ApJ, 747,
  L37

\bibitem[{{Majewski} {et~al.}(2003){Majewski}, {Skrutskie}, {Weinberg}, \&
  {Ostheimer}}]{M03}
{Majewski}, S.~R., {Skrutskie}, M.~F., {Weinberg}, M.~D., \& {Ostheimer}, J.~C.
  2003, ApJ, 599, 1082

\bibitem[{{Marino} {et~al.}(2011){Marino}, {Milone}, {Piotto}, {Villanova},
  {Gratton}, {D'Antona}, {Anderson}, {Bedin}, {Bellini}, {Cassisi}, {Geisler},
  {Renzini}, \& {Zoccali}}]{M11}
{Marino}, A.~F., {Milone}, A.~P., {Piotto}, G., {et~al.} 2011, ApJ, 731, 64

\bibitem[{{Marino} {et~al.}(2014){Marino}, {Milone}, {Przybilla}, {Bergemann},
  {Lind}, {Asplund}, {Cassisi}, {Catelan}, {Casagrande}, {Valcarce}, {Bedin},
  {Cort{\'e}s}, {D'Antona}, {Jerjen}, {Piotto}, {Schlesinger}, {Zoccali}, \&
  {Angeloni}}]{M14}
{Marino}, A.~F., {Milone}, A.~P., {Przybilla}, N., {et~al.} 2014, MNRAS, 437,
  1609

\bibitem[{{Martell} {et~al.}(2011){Martell}, {Smolinski}, {Beers}, \&
  {Grebel}}]{Martell11}
{Martell}, S.~L., {Smolinski}, J.~P., {Beers}, T.~C., \& {Grebel}, E.~K. 2011,
  A\&A, 534, A136

\bibitem[{{Mart{\'{\i}}nez-Delgado} {et~al.}(2012){Mart{\'{\i}}nez-Delgado},
  {Romanowsky}, {Gabany}, {Annibali}, {Arnold}, {Fliri}, {Zibetti}, {van der
  Marel}, {Rix}, {Chonis}, {Carballo-Bello}, {Aloisi}, {Macci{\`o}},
  {Gallego-Laborda}, {Brodie}, \& {Merrifield}}]{MD12}
{Mart{\'{\i}}nez-Delgado}, D., {Romanowsky}, A.~J., {Gabany}, R.~J., {et~al.}
  2012, ApJ, 748, L24

\bibitem[{{Mel{\'e}ndez} \& {Barbuy}(2002)}]{M02}
{Mel{\'e}ndez}, J. \& {Barbuy}, B. 2002, ApJ, 575, 474

\bibitem[{{Mel{\'e}ndez} {et~al.}(2006){Mel{\'e}ndez}, {Shchukina},
  {Vasiljeva}, \& {Ram{\'{\i}}rez}}]{M06}
{Mel{\'e}ndez}, J., {Shchukina}, N.~G., {Vasiljeva}, I.~E., \&
  {Ram{\'{\i}}rez}, I. 2006, ApJ, 642, 1082

\bibitem[{{Meza} {et~al.}(2005){Meza}, {Navarro}, {Abadi}, \&
  {Steinmetz}}]{M05}
{Meza}, A., {Navarro}, J.~F., {Abadi}, M.~G., \& {Steinmetz}, M. 2005, MNRAS,
  359, 93

\bibitem[{{Milone}(2015)}]{M15}
{Milone}, A.~P. 2015, \mnras, 446, 1672

\bibitem[{{Monaco} {et~al.}(2005){Monaco}, {Bellazzini}, {Bonifacio},
  {Ferraro}, {Marconi}, {Pancino}, {Sbordone}, \& {Zaggia}}]{Monaco05}
{Monaco}, L., {Bellazzini}, M., {Bonifacio}, P., {et~al.} 2005, \aap, 441, 141

\bibitem[{{Mucciarelli} {et~al.}(2014){Mucciarelli}, {Lovisi}, {Lanzoni}, \&
  {Ferraro}}]{Mucciarelli14}
{Mucciarelli}, A., {Lovisi}, L., {Lanzoni}, B., \& {Ferraro}, F.~R. 2014, \apj,
  786, 14

\bibitem[{{Navarro} {et~al.}(2004){Navarro}, {Helmi}, \& {Freeman}}]{NHF04}
{Navarro}, J.~F., {Helmi}, A., \& {Freeman}, K.~C. 2004, ApJ, 601, L43

\bibitem[{{Neves} {et~al.}(2009){Neves}, {Santos}, {Sousa}, {Correia}, \&
  {Israelian}}]{N09}
{Neves}, V., {Santos}, N.~C., {Sousa}, S.~G., {Correia}, A.~C.~M., \&
  {Israelian}, G. 2009, A\&A, 497, 563

\bibitem[{{Nidever} {et~al.}(2002){Nidever}, {Marcy}, {Butler}, {Fischer}, \&
  {Vogt}}]{Nid02}
{Nidever}, D.~L., {Marcy}, G.~W., {Butler}, R.~P., {Fischer}, D.~A., \& {Vogt},
  S.~S. 2002, ApJs, 141, 503

\bibitem[{{Nissen} {et~al.}(2002){Nissen}, {Primas}, {Asplund}, \&
  {Lambert}}]{N02}
{Nissen}, P.~E., {Primas}, F., {Asplund}, M., \& {Lambert}, D.~L. 2002, A\&A,
  390, 235

\bibitem[{{Norris}(2004)}]{N04}
{Norris}, J.~E. 2004, ApJ, 612, L25

\bibitem[{{Norris} \& {Da Costa}(1995)}]{N95}
{Norris}, J.~E. \& {Da Costa}, G.~S. 1995, ApJ, 447, 680

\bibitem[{{Obrien} \& {Lambert}(1986)}]{O86}
{Obrien}, Jr., G.~T. \& {Lambert}, D.~L. 1986, ApJs, 62, 899

\bibitem[{{Pasquini} {et~al.}(2011){Pasquini}, {Mauas}, {K{\"a}ufl}, \&
  {Cacciari}}]{P11}
{Pasquini}, L., {Mauas}, P., {K{\"a}ufl}, H.~U., \& {Cacciari}, C. 2011, A\&A,
  531, A35

\bibitem[{{Peck} \& {Reeder}(1972)}]{P72}
{Peck}, E.~R. \& {Reeder}, K. 1972, J. Opt. Soc. Am., 62, 958

\bibitem[{{Pereira} {et~al.}(2012){Pereira}, {Jilinski}, {Drake}, {de Castro},
  {Ortega}, {Chavero}, \& {Roig}}]{P12}
{Pereira}, C.~B., {Jilinski}, E., {Drake}, N.~A., {et~al.} 2012, A\&A, 543, A58

\bibitem[{{Piotto} {et~al.}(2007){Piotto}, {Bedin}, {Anderson}, {King},
  {Cassisi}, {Milone}, {Villanova}, {Pietrinferni}, \& {Renzini}}]{P07}
{Piotto}, G., {Bedin}, L.~R., {Anderson}, J., {et~al.} 2007, \apjl, 661, L53

\bibitem[{{Piotto} {et~al.}(2005){Piotto}, {Villanova}, {Bedin}, {Gratton},
  {Cassisi}, {Momany}, {Recio-Blanco}, {Lucatello}, {Anderson}, {King},
  {Pietrinferni}, \& {Carraro}}]{P05}
{Piotto}, G., {Villanova}, S., {Bedin}, L.~R., {et~al.} 2005, ApJ, 621, 777

\bibitem[{{Ram{\'{\i}}rez} {et~al.}(2007){Ram{\'{\i}}rez}, {Allende Prieto}, \&
  {Lambert}}]{R07}
{Ram{\'{\i}}rez}, I., {Allende Prieto}, C., \& {Lambert}, D.~L. 2007, A\&A,
  465, 271

\bibitem[{{Ram{\'{\i}}rez} {et~al.}(2013){Ram{\'{\i}}rez}, {Allende Prieto}, \&
  {Lambert}}]{R13}
{Ram{\'{\i}}rez}, I., {Allende Prieto}, C., \& {Lambert}, D.~L. 2013, ApJ, 764,
  78

\bibitem[{{Ram{\'{\i}}rez} {et~al.}(2014){Ram{\'{\i}}rez}, {Bajkova},
  {Bobylev}, {Roederer}, {Lambert}, {Endl}, {Cochran}, {MacQueen}, \&
  {Wittenmyer}}]{R14}
{Ram{\'{\i}}rez}, I., {Bajkova}, A.~T., {Bobylev}, V.~V., {et~al.} 2014, ApJ,
  787, 154

\bibitem[{{Ram{\'{\i}}rez} {et~al.}(2012){Ram{\'{\i}}rez}, {Mel{\'e}ndez}, \&
  {Chanam{\'e}}}]{R12}
{Ram{\'{\i}}rez}, I., {Mel{\'e}ndez}, J., \& {Chanam{\'e}}, J. 2012, ApJ, 757,
  164

\bibitem[{{Ram{\'{\i}}rez} {et~al.}(2011){Ram{\'{\i}}rez}, {Mel{\'e}ndez},
  {Cornejo}, {Roederer}, \& {Fish}}]{R11}
{Ram{\'{\i}}rez}, I., {Mel{\'e}ndez}, J., {Cornejo}, D., {Roederer}, I.~U., \&
  {Fish}, J.~R. 2011, ApJ, 740, 76

\bibitem[{{Reddy} {et~al.}(2006){Reddy}, {Lambert}, \& {Allende Prieto}}]{R06}
{Reddy}, B.~E., {Lambert}, D.~L., \& {Allende Prieto}, C. 2006, MNRAS, 367,
  1329

\bibitem[{{Reddy} {et~al.}(2003){Reddy}, {Tomkin}, {Lambert}, \& {Allende
  Prieto}}]{R03}
{Reddy}, B.~E., {Tomkin}, J., {Lambert}, D.~L., \& {Allende Prieto}, C. 2003,
  MNRAS, 340, 304

\bibitem[{{Renson} \& {Manfroid}(2009)}]{R09}
{Renson}, P. \& {Manfroid}, J. 2009, \aap, 498, 961

\bibitem[{{Sanz-Forcada} \& {Dupree}(2008)}]{S08}
{Sanz-Forcada}, J. \& {Dupree}, A.~K. 2008, A\&A, 488, 715

\bibitem[{{Sbordone} {et~al.}(2007){Sbordone}, {Bonifacio}, {Buonanno},
  {Marconi}, {Monaco}, \& {Zaggia}}]{S07}
{Sbordone}, L., {Bonifacio}, P., {Buonanno}, R., {et~al.} 2007, \aap, 465, 815

\bibitem[{{Searle} \& {Zinn}(1978)}]{S78}
{Searle}, L. \& {Zinn}, R. 1978, ApJ, 225, 357

\bibitem[{{Shetrone} {et~al.}(2003){Shetrone}, {Venn}, {Tolstoy}, {Primas},
  {Hill}, \& {Kaufer}}]{S03}
{Shetrone}, M., {Venn}, K.~A., {Tolstoy}, E., {et~al.} 2003, \aj, 125, 684

\bibitem[{{Smith} {et~al.}(2014){Smith}, {Dupree}, \& {Strader}}]{Smith14}
{Smith}, G., {Dupree}, A., \& {Strader}, J. 2014, ArXiv e-prints

\bibitem[{{Smith} {et~al.}(2012){Smith}, {Dupree}, \& {Strader}}]{S12}
{Smith}, G.~H., {Dupree}, A.~K., \& {Strader}, J. 2012, PASP, 124, 1252

\bibitem[{{Sneden} \& {Primas}(2001)}]{S01}
{Sneden}, C. \& {Primas}, F. 2001, NewA Rev., 45, 513

\bibitem[{{Soubiran} {et~al.}(2013){Soubiran}, {Jasniewicz}, {Chemin}, {Crifo},
  {Udry}, {Hestroffer}, \& {Katz}}]{S13}
{Soubiran}, C., {Jasniewicz}, G., {Chemin}, L., {et~al.} 2013, A\&A, 552, A64

\bibitem[{{Soubiran} {et~al.}(2010){Soubiran}, {Le Campion}, {Cayrel de
  Strobel}, \& {Caillo}}]{S10}
{Soubiran}, C., {Le Campion}, J.-F., {Cayrel de Strobel}, G., \& {Caillo}, A.
  2010, A\&A, 515, A111

\bibitem[{{Takeda} \& {Takada-Hidai}(2011)}]{T11}
{Takeda}, Y. \& {Takada-Hidai}, M. 2011, PASJ, 63, 547

\bibitem[{{Tolstoy} {et~al.}(2009){Tolstoy}, {Hill}, \& {Tosi}}]{T09}
{Tolstoy}, E., {Hill}, V., \& {Tosi}, M. 2009, \araa, 47, 371

\bibitem[{{Uttenthaler} {et~al.}(2011){Uttenthaler}, {van Stiphout}, {Voet},
  {van Winckel}, {van Eck}, {Jorissen}, {Kerschbaum}, {Raskin}, {Prins},
  {Pessemier}, {Waelkens}, {Fr{\'e}mat}, {Hensberge}, {Dumortier}, \&
  {Lehmann}}]{U11}
{Uttenthaler}, S., {van Stiphout}, K., {Voet}, K., {et~al.} 2011, A\&A, 531,
  A88

\bibitem[{{Valcarce} {et~al.}(2014){Valcarce}, {Catelan},
  {Alonso-Garc{\'{\i}}a}, {Cort{\'e}s}, \& {De Medeiros}}]{V14}
{Valcarce}, A.~A.~R., {Catelan}, M., {Alonso-Garc{\'{\i}}a}, J., {Cort{\'e}s},
  C., \& {De Medeiros}, J.~R. 2014, \apj, 782, 85

\bibitem[{{van Leeuwen}(2007)}]{vL07}
{van Leeuwen}, F. 2007, A\&A, 474, 653

\bibitem[{{Venn} {et~al.}(2004){Venn}, {Irwin}, {Shetrone}, {Tout}, {Hill}, \&
  {Tolstoy}}]{V04}
{Venn}, K.~A., {Irwin}, M., {Shetrone}, M.~D., {et~al.} 2004, AJ, 128, 1177

\bibitem[{{Ventura} {et~al.}(2001){Ventura}, {D'Antona}, {Mazzitelli}, \&
  {Gratton}}]{Ventura01}
{Ventura}, P., {D'Antona}, F., {Mazzitelli}, I., \& {Gratton}, R. 2001, ApJ,
  550, L65

\bibitem[{{Vesperini} {et~al.}(2010){Vesperini}, {McMillan}, {D'Antona}, \&
  {D'Ercole}}]{V10}
{Vesperini}, E., {McMillan}, S.~L.~W., {D'Antona}, F., \& {D'Ercole}, A. 2010,
  ApJ, 718, L112

\bibitem[{{Villanova} {et~al.}(2012){Villanova}, {Geisler}, {Piotto}, \&
  {Gratton}}]{V12}
{Villanova}, S., {Geisler}, D., {Piotto}, G., \& {Gratton}, R.~G. 2012, ApJ,
  748, 62

\bibitem[{{Villanova} {et~al.}(2009){Villanova}, {Piotto}, \& {Gratton}}]{V09}
{Villanova}, S., {Piotto}, G., \& {Gratton}, R.~G. 2009, A\&A, 499, 755

\bibitem[{{Vivas} {et~al.}(2001){Vivas}, {Zinn}, {Andrews}, {Bailyn}, {Baltay},
  {Coppi}, {Ellman}, {Girard}, {Rabinowitz}, {Schaefer}, {Shin}, {Snyder},
  {Sofia}, {van Altena}, {Abad}, {Bongiovanni}, {Brice{\~n}o}, {Bruzual},
  {Della Prugna}, {Herrera}, {Magris}, {Mateu}, {Pacheco}, {S{\'a}nchez},
  {S{\'a}nchez}, {Schenner}, {Stock}, {Vicente}, {Vieira}, {Ferr{\'{\i}}n},
  {Hernandez}, {Gebhard}, {Honeycutt}, {Mufson}, {Musser}, \&
  {Rengstorf}}]{V01}
{Vivas}, A.~K., {Zinn}, R., {Andrews}, P., {et~al.} 2001, ApJ, 554, L33

\bibitem[{{Wylie-de Boer} {et~al.}(2010){Wylie-de Boer}, {Freeman}, \&
  {Williams}}]{W10}
{Wylie-de Boer}, E., {Freeman}, K., \& {Williams}, M. 2010, AJ, 139, 636

\end{thebibliography}

\appendix

Figure~\ref{hr1996} shows the wavelength-calibrated spectrum of the hot, fast rotating HR~1996. 
Figures~\ref{crires1}--\ref{crires4} show all our CRIRES rest-frame spectra in the region of interest.

\setcounter{figure}{0}
\renewcommand{\thefigure}{A\arabic{figure}}

\begin{figure*}[h!]
\centering
\includegraphics[width=0.75\linewidth]{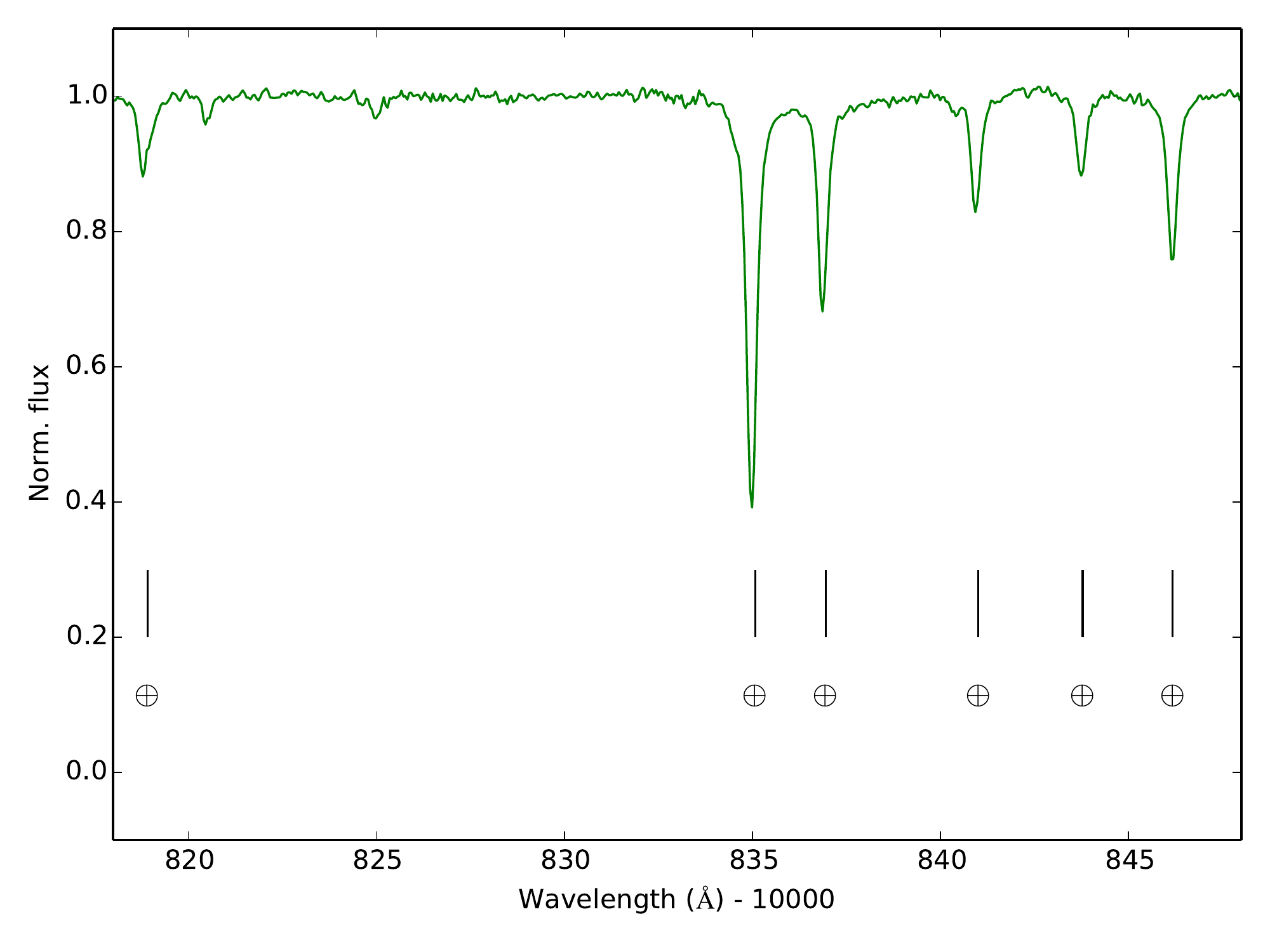}
\caption{CRIRES spectrum of HR 1996, a fast rotating star used as telluric standard. The most 
prominent water vapor absorption lines, in the (vacuum) rest frame, are marked with vertical lines. 
Wavelengths are taken from \cite{B73} and converted to vacuum using the formula of Peck \& 
Reeder (1972).}
\label{hr1996}
\end{figure*}

\begin{figure*}
\centering
\includegraphics[width=0.75\linewidth]{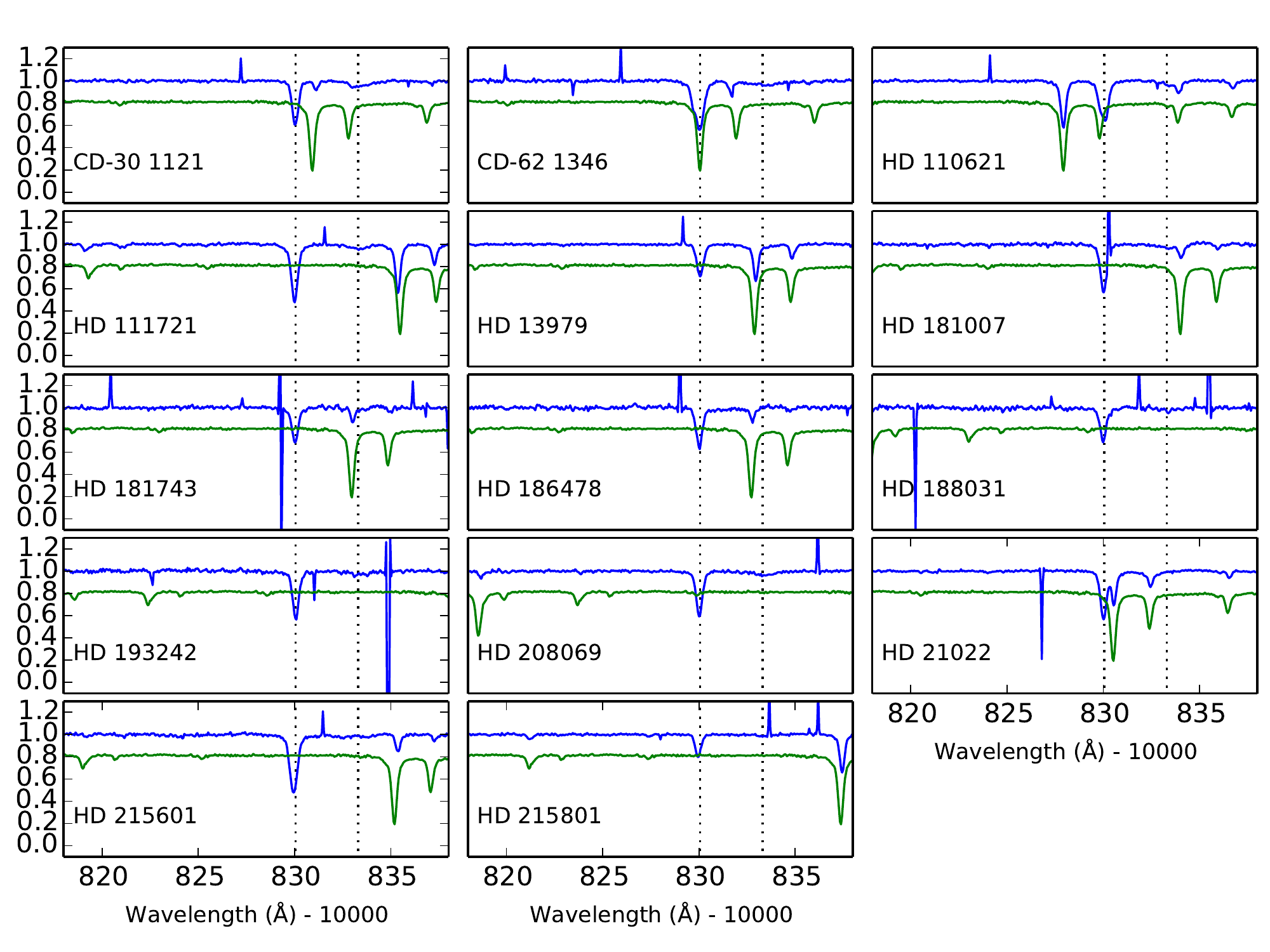}
\caption{CRIRES spectra for the Kapteyn group stars. In each panel, the top spectrum corresponds to the target stars and the bottom one 
is the hot fast rotating star HR~1996, which shows the telluric lines, shifted to match the wavelength rest-frame of the target star. 
The vertical dotted lines mark the position of the \sii\ line (left) and the \hei\ line (right).}
\label{crires1}
\end{figure*}
 
\begin{figure*}
\centering
\includegraphics[width=0.75\linewidth]{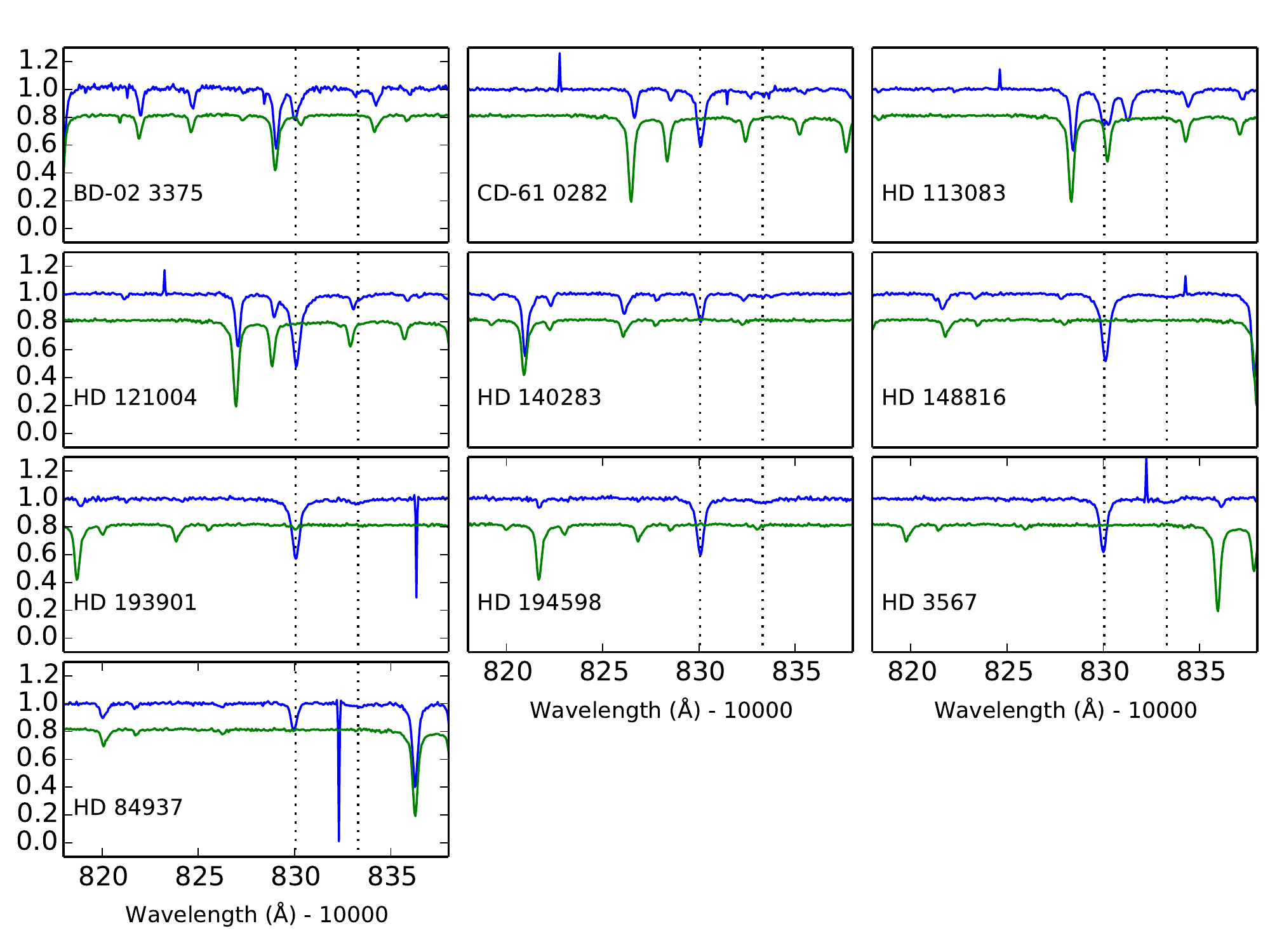}
\caption{CRIRES spectra for the $\omega$~Cen-group stars. 
The meaning of the two spectra in the panels as the vertical lines is the same as in Figure~\ref{crires1}.}
\label{crires2}
\end{figure*}
 
\begin{figure*}
\centering
\includegraphics[width=0.75\linewidth]{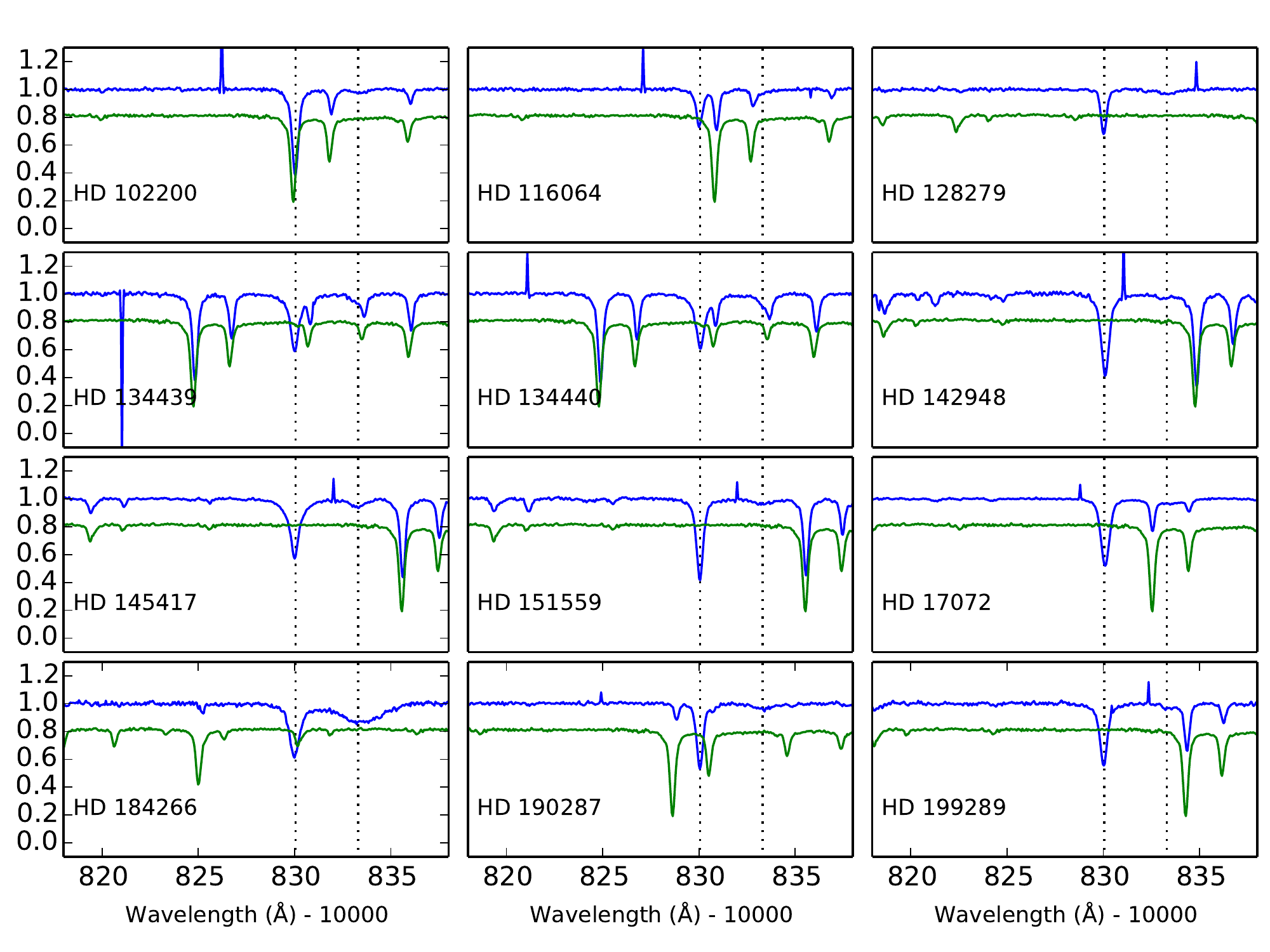}
\caption{CRIRES spectra for the field stars of our sample.}
\label{crires3}
\end{figure*}
 
\begin{figure*}
\centering
\includegraphics[width=0.75\linewidth]{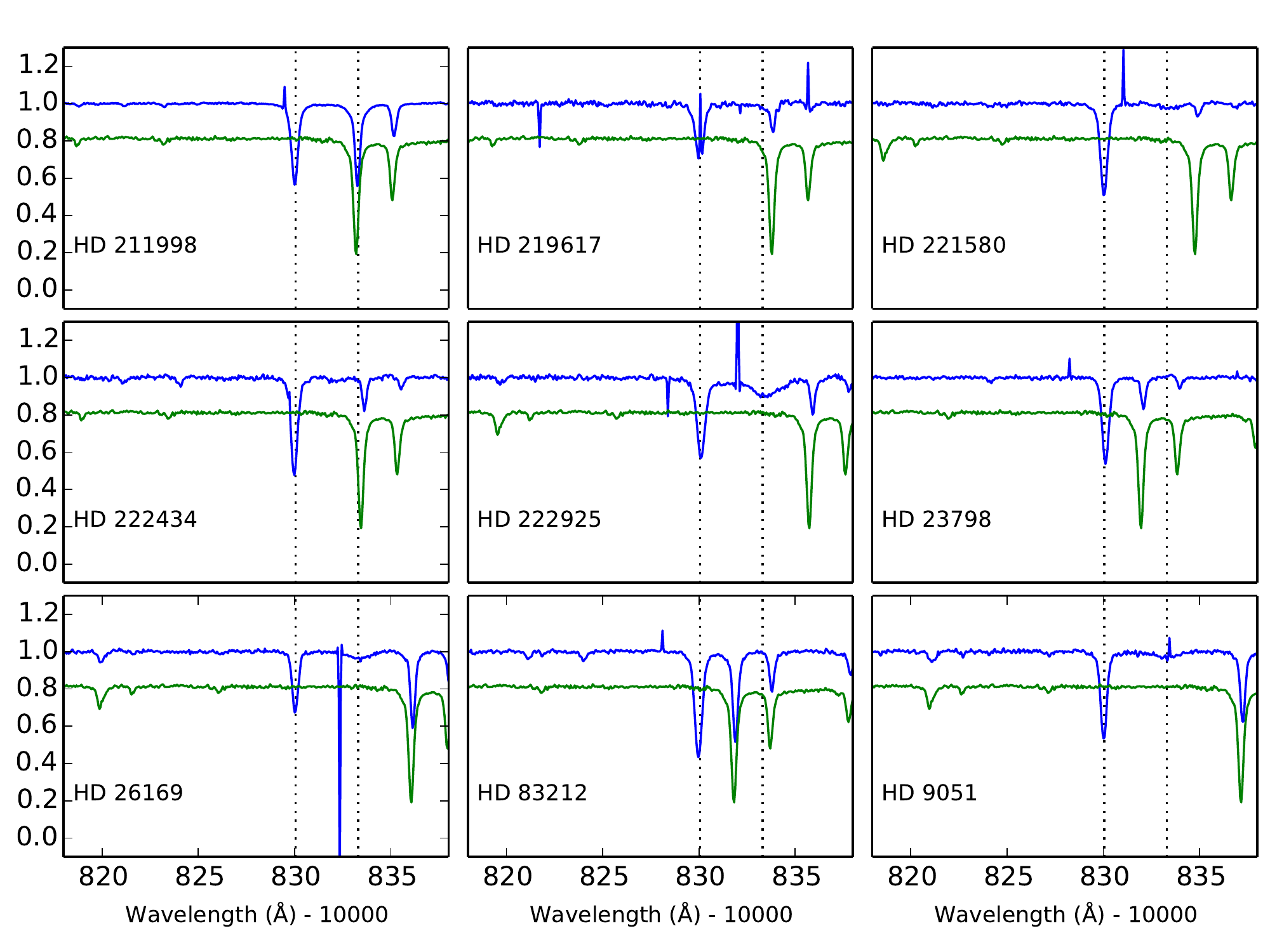}
\caption{CRIRES spectra for the field stars of our sample - continuation.}
\label{crires4}
\end{figure*}

\end{document}